\documentclass[aps,amsmath,amssymb,twocolumn]{revtex4-1}

\usepackage{graphicx}
\usepackage{dcolumn}
\usepackage{bm}
\usepackage{color}

\begin{document}

\title{The nature of the electrical conduction in ferromagnetic atomic-size contacts:\\
insights from shot noise measurements and theoretical simulations}

\author{R. Vardimon$^{1}$}
\author{M. Matt$^{2}$}
\author{P. Nielaba$^{2}$}
\author{J. C. Cuevas$^{3}$}
\author{O. Tal$^{1}$}

\affiliation{$^{1}$Department of Chemical Physics, Weizmann Institute of Science, Rehovot 76100, Israel}

\affiliation{$^{2}$Department of Physics, University of Konstanz, D-78457 Konstanz, Germany}

\affiliation{$^{3}$Departamento de F\'{\i}sica Te\'orica de la Materia Condensada and Condensed Matter
Physics Center (IFIMAC), Universidad Aut\'onoma de Madrid, E-28049 Madrid, Spain}

\date{\today}

\begin{abstract}
With the goal to elucidate the nature of spin-dependent electronic transport in ferromagnetic atomic contacts, we 
present here a combined experimental and theoretical study of the conductance and shot noise of metallic atomic 
contacts made of the $3d$ ferromagnetic materials Fe, Co, and Ni. For comparison, we also present the corresponding 
results for the noble metal Cu. Conductance and shot noise measurements, performed using a low-temperature break 
junction setup, show that in these ferromagnetic nanowires: (i) there is no conductance quantization of any kind, 
(ii) transport is dominated by several partially-open conduction channels, even in the case of single-atom contacts, 
and (iii) the Fano factor of large contacts saturates to values that clearly differs from those of monovalent 
(nonmagnetic) metals. We rationalize these observations with the help of a theoretical approach that combines 
molecular dynamics simulations to describe the junction formation with nonequilibrium Green's function techniques 
to compute the transport properties within the Landauer-B\"uttiker framework. Our theoretical approach successfully 
reproduces all the basic experimental results and it shows that all the observations can be traced back to the fact 
that the $d$ bands of the minority-spin electrons play a fundamental role in the transport through ferromagnetic 
atomic-size contacts. These $d$ bands give rise to partially open conduction channels for any contact size, which 
in turn lead naturally to the different observations described above. Thus, the transport picture for these nanoscale 
ferromagnetic wires that emerges from the ensemble of our results is clearly at variance with the well established 
conduction mechanism that governs the transport in macroscopic ferromagnetic wires, where the $d$ bands are 
responsible for the magnetism but do not take part in the charge flow. These insights provide a fundamental 
framework for ferromagnetic-based spintronics at the nanoscale.
\end{abstract}

\pacs{73.40.Jn, 73.63.Rt, 72.25.Ba, 75.76.+j}


\maketitle

\section{Introduction}

When the characteristic dimensions of a metallic wire are shrunk all the way down to the atomic scale, its electronic transport 
properties change dramatically as a consequence of the appearance of quantum mechanical effects. This has been nicely illustrated 
in recent years with the help of metallic atomic contacts fabricated by means of scanning tunneling microscopes (STMs) and break-junction 
techniques \cite{Agrait2003}. Thus for instance, it has been shown that transport properties such as the conductance 
\cite{Agrait2003,Scheer1998}, shot noise \cite{vandenBrom1999,Wheeler2010,Chen2012,Kumar2012,Vardimon2013,Chen2014}, or thermopower 
\cite{Ludoph1999,Tsutsui2013,Evangeli2015}, differ markedly from those of macroscopic wires, while transport phenomena such as Joule 
heating \cite{Smit2004,Lee2013} or magnetoresistive effects \cite{Bolotin2006a,Sokolov2007,Neel2009,Egle2010,Strigl2015,Vardimon2015} 
take place in a very different manner in these nanowires. In all cases, these dramatic differences can be traced back to the fact 
that the transport in metallic atomic contacts is mainly quantum coherent. The central goal of this work is to show that the nature 
of electrical conduction in ferromagnetic atomic contacts is clearly at variance with the established picture in macroscopic wires.

Our present understanding of electrical conduction in macroscopic wires made of ferromagnetic metals like Fe, Co, and Ni is largely 
based on the semiclassical model put forward by Sir Nevill Mott in the 1930s \cite{Mott1936a,Mott1936b}. In that simple model, sometimes 
referred to as the two-current model, the current is carried by the electrons of two independent spin bands, the majority-spin 
electrons and the minority-spin electrons. Moreover, it is assumed that the conduction bands have an $s$ character, while the 
$d$-electrons are localized and the corresponding spin-split bands are responsible for the net magnetization. Mott described the 
conductivity of a ferromagnetic metal in terms of the Drude formula, where the key parameters are the density of conduction electrons, 
assumed to be spin-independent, and the spin-dependent inelastic scattering time (or relaxation time). While the relaxation time for 
the majority-spin electrons is similar to that of a nonmagnetic metal, the scattering of minority-spin electrons from the $s$ states 
into the partially occupied $d$ states reduces considerably the corresponding relaxation time and, in turn, the electrical conductivity 
in ferromagnetic metals. This explains, for instance, why Fe, Co, and Ni are less conductive than the noble metals (Au, Ag, and Cu). 
Moreover, this model predicts that the conduction is dominated by the majority-spin electrons, leading to a positive spin polarization 
of the current. This simple picture turns out to be quite accurate, as it has been very recently demonstrated using ultrafast terahertz 
spectroscopy \cite{Jin2015}. However, as we shall show in this work, it severely fails to explain the transport properties in ferromagnetic 
atomic contacts.

Since the advent of STM-based and break-junction techniques in the 1990s, many experimental studies of transport through atomic 
contacts made of ferromagnetic metals have been reported
\cite{Sirvent1996,Costa1997,Hansen1997,Ott1998,Oshima1998,Komori1999,Ono1999,Garcia1999,Ludoph2000b,
Yanson2001,Viret2002,Elhoussine2002,Shimizu2002,Gillingham2002,Rodrigues2003,Gillingham2003a,Gillingham2003b,Untiedt2004,Gabureac2004,
Yang2004,Costa2005,Bolotin2006b,Keane2006,Bolotin2006a,Viret2006,Sokolov2007,Neel2009,Calvo2009,Halbritter2010,Egle2010,Moriguchi2012,
vonBieren2013,Strigl2015,Vardimon2015}. Some of the early reports focused on the observation of half-integer quantization, \emph{i.e.}, 
on the observation of peaks in the conductance histograms at half-integer multiples of the quantum of conductance $G_0 = 2e^2/h$ 
\cite{Elhoussine2002,Shimizu2002,Gillingham2002,Rodrigues2003,Gillingham2003a,Gillingham2003b,Sokolov2007}. In the Landauer picture of 
the coherent transport through these ferromagnetic nanowires, the low-temperature linear conductance is given by $G=(G_0/2) \sum_{n,\sigma} 
\tau_{n,\sigma}$, where $\tau_{n,\sigma}$ is the transmission coefficient at the Fermi energy of the $n$th available conduction channel 
for a spin $\sigma$. Thus, these observations were interpreted as an evidence indicating that only spin-split fully open channels 
contribute to the conductance in these atomic-scale ferromagnetic wires. This interpretation was questioned by Untiedt \emph{et al.}\ 
\cite{Untiedt2004} who measured the conductance for atomic contacts of Fe, Co, and Ni using break junctions under cryogenic vacuum 
conditions. Contrary to the experiments mentioned above, they reported the absence of fractional conductance quantization. Instead, 
they observed conductance histograms that show broad peaks above $1G_0$. Moreover, they suggested that the observation of peaks in 
the conductance histograms at half-integer values of $G_0$ could be due to contamination. We note that half-integer conductance 
values are occasionally detected for arbitrarily realized atomic-scale contacts of both ferromagnetic and non-ferromagnetic metals.

Another controversy in the context of ferromagnetic atomic contacts has to do with the observation of an anomalous anisotropic 
magnetoresistance (AMR). AMR is a spintronic effect in which the resistance varies as a function of the relative orientation between 
the magnetization and the current directions and it originates from spin-orbit interaction. Bolotin \emph{et al.}\ \cite{Bolotin2006a} 
reported that the magnitude of the AMR of permalloy atomic contacts can be considerably larger than in bulk samples and that it exhibits 
an anomalous angular dependence. Similar observations were reported by Viret \emph{et al.}\ \cite{Viret2006} in Fe contacts, but 
they also reported the occurrence of conductance jumps upon the rotation of the magnetization. Similar step-wise variations in the 
conductance were later found in Co nanocontacts \cite{Sokolov2007}. These jumps were tentatively interpreted as a manifestation of 
the so-called ballistic AMR (BAMR) \cite{Velev2005}. This approach suggested that in a ballistic contact the rotation of the 
magnetization could result in variations in band crossing at the Fermi energy, leading to an abrupt change in the conductance on 
the order of $G_0/2$. However, as discussed above, ferromagnetic contacts are not expected to be ballistic (\emph{i.e.}\, to exhibit 
only fully-open channels), and therefore, the interpretation in terms of BAMR is highly questionable \cite{Hafner2009}. In fact, 
Shi and Ralph \cite{Shi2007a} suggested that these jumps might originate from sudden atomic rearrangements \cite{Shi2007b}.

Half-integer quantization and BAMR are not inherently expected in ferromagnetic atomic contacts in view of the established picture 
of the conduction in these atomic-scale wires \cite{Scheer1998,Cuevas1998}. Within this picture, based on the Landauer approach 
to coherent transport, the conduction channels in a nonmagnetic metal are spin-degenerate and are determined by the orbital 
structure of the metal and the geometry of the atomic contact. In particular, the number of conduction channels of a single-atom 
contact is expected to be limited by the number of valence orbitals, as it was experimentally verified with the use of 
superconductivity to extract the channel content \cite{Scheer1998}. Moreover, it was established that conductance quantization 
is only expected in few-atom contacts of monovalent metals, whereas in the case of multivalent metals the conductance is essentially  
determined by several partially-open channels and there are no signs of conductance quantization.

The controversies discussed above and the great interest in spin-dependent transport at the nanoscale have led to numerous 
theoretical groups to investigate the transport properties of ferromagnetic atomic contacts \cite{Martin2001,Krstic2002,Smogunov2002,Delin2003,Velev2004,Rocha2004,Bagrets2004,Jacob2005,Wierzbowska2005,
Dalgleish2005,Khomyakov2005,Fernandez-Rossier2005,Velev2005,Smogunov2006,Pauly2006,Jacob2006,Autes2006,Xia2006,
Rocha2007,Tung2007,Bagrets2007,Autes2008,Jacob2008,Hafner2009,Tao2010,Hardrat2012,Xie2012,Calvo2012,Tan2013}.
Apart from the analysis of the conductance and the AMR, these studies have also addressed many different aspects of the physics 
of these contacts such as the electronic structure of ideal systems, like monoatomic wires, the influence of domain walls on 
electronic and transport properties, or the magnetic structure. The ensemble of the reported theoretical results clearly suggest 
that the $d$ orbitals play a fundamental role in the transport properties of these atomic-scale wires, contrary to their macroscopic 
counterparts, and there is no fundamental reason for expecting either conductance quantization or full spin polarization. However, 
there is still no generally accepted picture for electrical conduction in ferromagnetic atomic contacts for two main reasons. 
First, a systematic one-to-one comparison between experiment and theory for the conductance has never been established. This is 
a difficult task since it requires from the theory side to obtain a simultaneous description of the detailed atomic structure 
and transport properties of the contact. Second and more important, transport measurements performed so far in ferromagnetic 
atomic contacts have been restricted to the conductance, which does not contain information about the individual transmission 
coefficients. The access to this information was the key breakthrough that finally enabled to elucidate the nature of the 
conduction in nonmagnetic atomic contacts, something that became possible thanks to the use of superconductivity 
\cite{Scheer1998,Scheer1997}. However, such a method is not possible in the case of ferromagnetic contacts.

In this work we revisit the nature of the electrical conduction in ferromagnetic atomic contacts by providing a very detailed 
experimental study of both conductance and shot noise in Fe, Co, and Ni atomic contacts at cryogenic temperature and vacuum 
conditions. We present the corresponding results for Cu contacts for comparison. In particular, we show how shot noise 
measurements provide an insight into the electrical conduction of these ferromagnetic nanowires, which is crucial to understand 
how they differ from nonmagnetic atomic-scale contacts and macroscopic ferromagnetic wires. Moreover, we supplement our 
experimental study with a comprehensive theoretical analysis based on molecular dynamics (MD) and quantum mechanical 
calculations of the transport properties that allows us to establish a direct comparison with our experiments for the different 
materials under study. From the experimental side, our main results for the ferromagnetic contacts are: (i) the confirmation 
of the lack of any kind of conductance quantization, (ii) we find no signs of pronounced shot noise suppression at any 
conductance value, contrary to noble metals, which shows that the transport is dominated by partially-open channels for 
any contact size, (iii) the analysis of shot noise shows that at least several channels contribute to transport even in smallest 
(presumably one-atom) contacts, again in clear contrast with noble metals, and (iv) the Fano factor saturates for contacts 
with conductances higher than $10G_0$ to a value that is clearly larger than for monovalent metals. Our 
theoretical results reproduce very satisfactorily all these basic observations. In all cases, the origin of these observations 
can be traced back to the fact that the $3d$ orbitals play a major role in the conduction of these ferromagnetic atomic contacts. 
The picture that emerges is that in a ferromagnetic contact the majority-spin electrons behave as in a noble metal, where the 
conduction is dominated by the $s$ conduction band, while the minority-spin electrons behave as in a transition metal, where 
both $s$ and $d$ orbitals contribute decisively to the transport properties. In particular, the $d$ orbitals corresponding 
to the minority-spin bands build up partially open channels which, in turn, are responsible for the experimental observations 
described above. Moreover, these additional conduction channels are responsible for the higher conductance of ferromagnetic 
few-atom contacts as compared with noble metals and they give rise to a negative spin polarization of the current in all 
three ferromagnetic $3d$ metals, in clear contrast to the positive spin polarization in macroscopic wires. Thus, our study 
provides a consistent general picture of the conduction in ferromagnetic atomic contacts and their unique transport properties 
compared to macroscopic ferromagnetic wires.

The rest of the manuscript is organized as follows. In section \ref{sec-Methods} we describe the experimental and theoretical 
techniques that we have used to investigate both the conductance and the shot noise of ferromagnetic atomic contacts. We present 
in section \ref{sec-Histograms} our results for the conductance histograms of the different materials studied in this work. 
Section \ref{sec-Noise} is devoted to the description of our experimental and theoretical results for the shot noise and Fano 
factor. We present a detailed discussion on the origin of our results in terms of the nature of the conduction channels in 
section \ref{sec-Channels}. Section \ref{sec-further-discussions} is devoted to further discussion of our results. Then, we 
summarize our main conclusions in section \ref{sec-Conclusions}. Finally, in Appendix A we show the results for Co and Fe 
that are not presented in the main text.

\section{General methodology} \label{sec-Methods}

In this section we describe the basic experimental and theoretical methods that we have employed in this work to study both the 
conductance and shot noise of ferromagnetic atomic contacts.

\begin{figure}[b]
\begin{center} \includegraphics[width=0.95\columnwidth,clip]{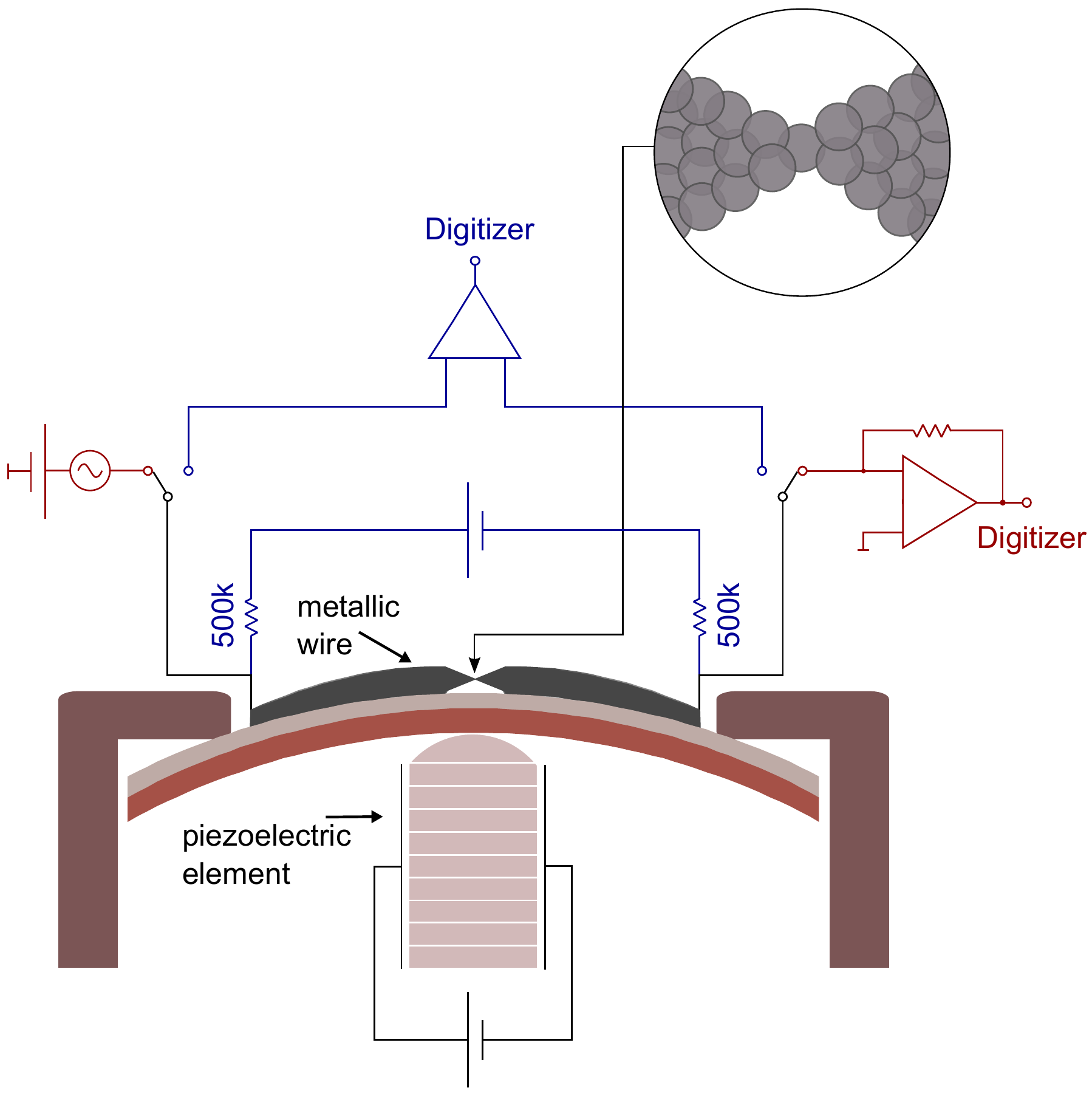} \end{center}
\caption{(Color online) Schematics of a mechanical controllable break junction setup adjusted for shot noise measurements. A 
notched metallic wire is broken in a controllable fashion using a piezoelectric element, allowing the formation of an atomic 
contact (inset). The electronic circuit enables switching between conductance (red) and shot noise (blue) measurement modes.}
\label{exp-setup}
\end{figure}

\subsection{Experimental techniques}

Our measurements were performed using a mechanical controllable break junction (MCBJ) \cite{Muller1992} composed of a notched 
metal wire attached to a flexible substrate (see Fig.~\ref{exp-setup}). The wires used consist of a high purity metal 
($\ge$ 99.9\%) and have macroscopic dimensions (0.1 mm diameter). The sample is positioned in a vacuum chamber that is 
initially pumped to $\sim 10^{-5}$ mbar and cooled by liquid helium to $\sim 4.2$ K. Once the sample is cold, a mechanical 
push screw is used to bend the substrate until the wire breaks in its weakest spot, forming two atomically-sharp tips. Since 
the junction is held under cryogenic vacuum, the freshly exposed tips are kept with minimum exposure to impurities. This is 
extremely important to avoid oxidation or other contaminations, which can have a considerable effect on transport 
\cite{Untiedt2004,Vardimon2015}. The displacement between the tips is then controlled with sub-Angstrom resolution by the 
precise movement of a piezoelectric element. The tips can be pushed back towards each other to reform an atomic contact, 
whose minimal cross section can be varied by changing the voltage applied on the piezo element.

The electronic conductance across the atomic contact is measured by connecting measurement wires to the two sides of the 
thinned metallic wire, which essentially serve as leads to the atomic-scale junction. Figure~\ref{exp-setup} shows the 
electronic circuit used for conductance and shot noise measurements. Two computer-controlled relays are used to toggle 
between conductance (marked in red) and noise (blue) measurement modes. For DC conductance measurement, a bias voltage 
is provided to the junction from a differential voltage source. The response current is amplified with a current preamplifier 
(SR 570) and recorded by a 24bit digitizer (NI 4461). AC conductance measurements are performed by driving a small sinusoidal 
signal (2 mVpp, $\sim$3 kHz) added to the bias voltage. The differential conductance $dI/dV$ is obtained using the lock-in 
technique, performed digitally with the data acquisition software.

For noise measurements, the sample is current biased and the voltage response is amplified by low-noise voltage amplifiers. 
The amplified voltage signal is recorded by a fast digitizer (NI 5992) and the power spectrum is calculated by digital 
Fourier transform. Two different configurations of voltage amplifiers were used for the experiments. In one configuration, 
a NF Li-75a was used, followed by a Signal Recovery 5184. The overall voltage amplification is $10^5$ and the input voltage 
noise is 1.5nV/$\sqrt{\mbox{Hz}}$. In a second configuration, a specially designed amplifier (JanasCard) with amplification 
of $10^4$ and voltage noise of 0.9nV/$\sqrt{\mbox{Hz}}$ is used. The lower voltage noise enabled a higher signal to noise 
ratio, which was important for shot noise measurements at high conductance values ($G=5$-$15G_0$). The sample and amplifiers 
are located inside a specially-designed Faraday cage in order to minimize noise pickup from environmental radiation. The 
piezo voltage is supplied by a differential voltage source and amplified by a factor of four using a Piezomechanik SVR-150 
piezo driver.

\subsection{Theoretical approach} \label{sec-methods-thy}

In order to compute the conductance and shot noise of the atomic contacts studied
experimentally, we have combined classical molecular dynamics (MD) simulations of the formation of the contacts, a tight-binding 
description of the electronic structure, and nonequilibrium Green's function techniques. Our methodology proceeds along the lines 
of Refs.~\cite{Dreher2005,Pauly2006,Pauly2011,Schirm2013,Chen2014,Evangeli2015}. In the following we briefly describe our approach.
\begin{figure}[t]
\begin{center} \includegraphics[width=\columnwidth,clip]{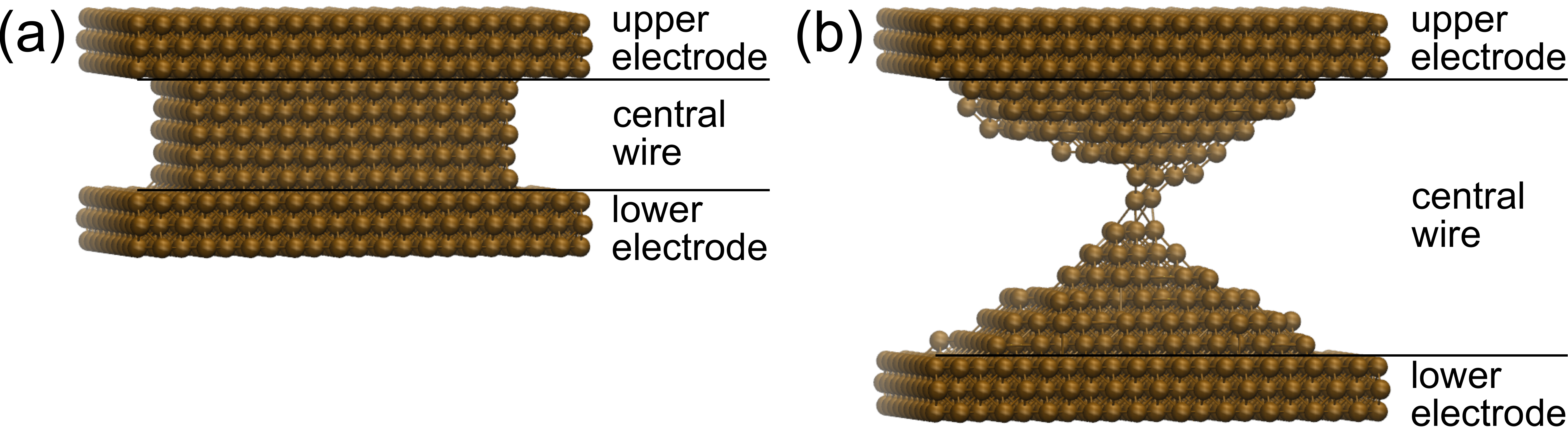} \end{center}
\caption{(Color online) Contact geometries in the molecular dynamics simulations. (a) Ideal fcc initial
structure of the Ni contacts. (b) The same Ni contact as in panel (a) after an elongation of $1.4$ nm.
In both panels we have indicated the partitioning of the contact into the upper and lower electrodes and
the central wire, as used for the MD and transport calculations.}
\label{fig-structure}
\end{figure}

\emph{Molecular dynamics simulations}. In metallic atomic contacts there is a crucial interplay between mechanical and transport 
properties. Thus, in order to establish a direct comparison with our experiments, it is necessary to describe the formation process 
of these nanowires. For this purpose we have carried out MD simulations using the open source program package LAMMPS 
\cite{Plimpton1995,LAMMPS-web}. Within LAMMPS, we employed the embedded atom method with the semi-empirical potentials from 
Ref.~\cite{Sheng2011} for Ni and Cu, Ref.~\cite{Purja2012} for Co, and Ref.~\cite{Mendelev2003} for Fe to model the interactions 
between the atoms. These potentials take into account the possibility to have an atomic coordination different from bulk. To 
generate the geometrical configurations, we started with an ideal fcc lattice for Co, Ni, and Cu and an ideal bcc lattice for 
Fe where the crystal direction $\langle 100\rangle$ lies parallel to the $z$ axis, coinciding with the transport and elongation 
direction. For the simulations, we first divided the geometry into three parts: Two electrodes and a central wire, attached 
to them (see Fig.~\ref{fig-structure}). The electrodes consist of 661 (321) atoms for fcc (bcc) each that are kept fixed 
during the MD calculations. The wire is made up of 563 (275) atoms that follow the Newtonian equations of motion. We assume a 
canonical ensemble and use the velocity Verlet integration scheme \cite{Frenkel2004}. The simulated wires have an initial 
length of 0.57 nm for Fe, 0.75 nm for Co, and 0.73 nm for Ni and Cu. The starting velocities of the atoms in the wire were 
chosen randomly with a Gaussian distribution to yield an average temperature of $T = 4$ K. Because of this randomness, every 
elongation calculation evolves differently, while a Nos\'e-Hoover thermostat keeps the temperature fixed \cite{Frenkel2004}. 
To relax the system, the wire gets equilibrated for 0.1 ns at the beginning of the elongation process. Finally, the elongation 
process is simulated by separating one electrode from the other at a constant velocity of 0.4 m/s. During this process, every 
10 ps the geometry is recorded. A stretching process needs a total simulation time of about 5 ns, until the contact breaks.

\emph{Transport calculations}. Once the geometries of the atomic wires are determined through the MD simulations, we use them to 
calculate the conductance and the shot noise within the Landauer-B\"uttiker formalism. Within this formalism all linear response 
transport properties are determined by the transmission function. To compute this function one needs, first of all, a proper 
description of the electronic structure of the metallic atomic contacts. For this purpose, we have employed the non-orthogonal 
Slater-Koster tight-binding parameterization of Refs.~\cite{Mehl1996,Mehl1998}, which has been quite successful in determining 
a variety of properties of these atomic wires \cite{Dreher2005,Pauly2006,Pauly2011,Schirm2013,Chen2014,Evangeli2015}
and it is known to accurately reproduce the band structure and total energy of bulk ferromagnetic materials \cite{Bacalis2001}. 
In this parameterization one takes into account the relevant valence orbitals of the material under study. For the four materials 
studied in this work (Fe, Co, Ni, and Cu), the atomic basis includes the $3d$, $4s$, and $4p$ orbitals. For the ferromagnetic 
materials, the on-site energies and hopping matrix elements depend on the electron spin and the model describes two independent 
sets of spin bands. In other words, there is no spin mixing in our model, as in the classical two-current model, which means 
in particular that we do not consider the spin-orbit interaction. This also means that we do not consider here the possibility 
of having magnetic domains in the atomic contacts. Within this tight-binding model, the hopping and overlap matrix elements 
are parametric functions of the distance between the atoms, which allows us to combine it with our MD simulations.

We compute the transmission in the framework of our tight-binding model by making use of nonequilibrium Green's function techniques, 
as described in Refs.~\cite{Dreher2005,Pauly2006,Pauly2011}. Briefly, as in the MD simulations, the system is divided into three regions 
for the transport calculations, \emph{i.e.}, the upper and lower electrodes and the central wire (see Fig.~\ref{fig-structure}). As 
the local environment of the atoms in the central part is very different from that in the bulk, we enforce the charge neutrality for 
all the atoms of the wire \cite{Pauly2006}. Such a neutrality condition is typically a good approximation for metallic systems. 
The electrodes are considered to be semi-infinite perfect crystals. Their surface Green's functions are computed with the help of a 
decimation technique \cite{Pauly2006,Guinea1983,Pauly2008}, and we use the same tight-binding parameterization as for the
central part to determine their electronic structure. It is worth stressing that the Green's function techniques allow us to compute 
not only the transmission function, but also the spin-resolved transmission eigenvalues, $\{\tau_{n,\sigma}\}$, the analysis of which 
provides additional physical insight. Within the Landauer-B\"uttiker formalism the conductance $G$ can be expressed in terms of 
$\{\tau_{n,\sigma} \}$ at the Fermi energy as \cite{Cuevas2010}
\begin{eqnarray}
\label{eq-G-S}
G & = &  \frac{G_0}{2} \sum_{n,\sigma} \tau_{n,\sigma} ,
\end{eqnarray}
where $\sigma=\uparrow, \downarrow$ is the spin. At low temperatures and within the linear regime, the zero-frequency shot 
noise is given by $S_I = 2eIF$, where $I$ is the bias current. The Fano factor $F$ describes the noise suppression with 
respect to its full Poissonian value of $2eI$, and is given by
\begin{equation}
\label{eq-Fano}
F = \frac{\sum_{n,\sigma} \tau_{n,\sigma}(1 - \tau_{n,\sigma})}{\sum_{n,\sigma} \tau_{n,\sigma}} .
\end{equation}
To conclude this section, let us introduce the concept of minimum cross section (MCS), which provides a measure of the contact 
size and the number of atoms in the narrowest part of the wire. As we shall see, this is a useful concept, but it must be
acknowledged that there is no unambiguous way to define this quantity in an atomic-scale wire. In our simulations, we
define and obtain this quantity as follows. We first superimpose our atomic structure with a fine-meshed grid with a grid
spacing of about 0.02 nm. In analogy to a sandglass, we compute the total flow through the contact, which consists of the 
flows through all (one or more) bottlenecks connecting the two ends of the wire. Dividing the total flow by the possible 
flow through one atom, one gets the sum of the areas of all bottlenecks as function of atomic cross sections.

\begin{figure*}[t]
\begin{center} \includegraphics*[width=0.85\textwidth,clip]{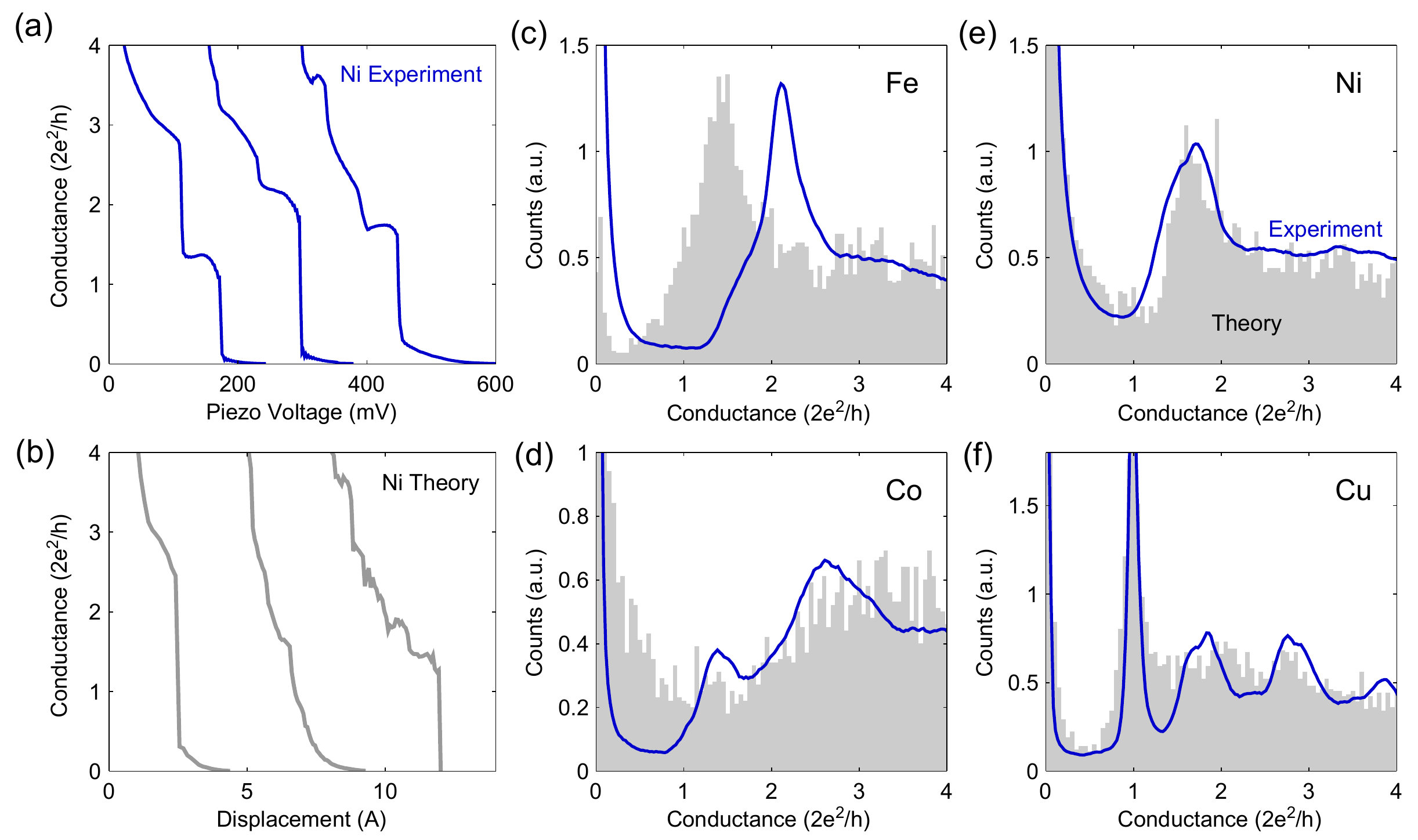} \end{center}
\caption{(Color online) (a) Ni conductance traces recorded in experiment at a bias of 50 mV. (b) Simulated Ni conductance traces. 
(c-f) Conductance histograms for Fe (c), Co (d), Ni (e) and Cu (f), constructed from experimental (blue line) traces, recorded 
at bias voltages of 50-150 mV, and simulated (gray area) traces. The experimental histograms were constructed from 10000 
traces for Co, Ni and Cu and 5000 traces for Fe. The theoretical histograms include 100 traces each.}
\label{cond-histo}
\end{figure*}

\section{Conductance histograms} \label{sec-Histograms}

To characterize the differences in transport and structural properties of Fe, Co, Ni and Cu atomic contacts,
we have measured their dc conductance ($I/V$) as function of applied piezo voltage, during repeated breaking and reforming 
of the contact. Figure~\ref{cond-histo}(a) shows examples of conductance traces recorded in the last stages of the breaking 
process of Ni contacts. As the minimal cross section of the contact is reduced, the conductance exhibits
a sequence of plateaus separated by sudden jumps to lower conductance values. The plateaus correspond to relatively stable
atomic configurations, which are separated by jumps due to sudden atomic rearrangements that occur once sufficient stress
is accumulated \cite{Rubio1996}. Further elongation leads to rupture of the atomic contact and to a corresponding drop
in the conductance. The small, though non-vanishing measured conductance arises from electron tunneling between the broken
atomic tips. Conductance traces obtained from our simulations of Ni contacts are presented in Fig.~\ref{cond-histo}(b).
As one can see, they nicely reproduce the slope of the conductance plateaus as well as the transition to the tunneling
regime.

The observed variation from trace to trace stems from the different atomic geometries probed during each breaking cycle. 
To obtain a statistical picture of the typical conductance values characterizing each metal, we construct conductance 
histograms by collecting conductance values from thousands of experimental conductance traces. Figures \ref{cond-histo}(c-f) 
show the conductance histograms obtained from experimental traces (blue lines). The histograms show different sets of 
peaks, which are interpreted as the conductance of frequently occurring atomic configurations \cite{Agrait2003}. For all 
ferromagnetic metals we find in common that the conductance peaks are located above the quantum of conductance ($G_0$) 
and that they exhibit a considerable width (FWHM $\sim 1G_0$). These observations are in good agreement with previously 
reported measurements in inert environments by Untiedt \emph{et al.} \cite{Untiedt2004}. In contrast, in the case of 
the monovalent metal Cu, the conductance histogram (Fig.~\ref{cond-histo}(f)) exhibits a series of relatively narrow 
peaks close to multiples of $G_0$. Our results therefore indicate the absence of any kind of conductance quantization 
(either integer or half-integer) for ferromagnetic contacts.

For comparison with the experimental data, we have computed conductance histograms from 100 breaking simulations for every metal 
(grey areas in Fig.~\ref{cond-histo}(c-f)). As one can see, there is generally a good agreement between theory and experiment 
in the locations and widths of the peak features for the different metals. One exception is the observed shift in the peak 
position of the simulated Fe histogram with respect to experiment. A possible explanation for this difference could be that 
the Fe potential, created for the crystalline phase with defects and for the liquid phase, is not able to provide a good 
enough description of the low-coordinated environment near the ending of the rupture process, despite its advantages from 
the embedded atom method (EAM) formalization. We tried different EAM potentials and different elongation directions, but 
the agreement with the experimental data did not improve significantly. Nevertheless, the reproduction of most of the 
experimental features indicates that our simulations provide a good description of the structural and transport properties 
of the examined atomic contacts. This is reinforced by the good agreement found for the shot noise results, as we shall 
discuss in the next section. 

\begin{figure}[t]
\begin{center} \includegraphics[width=\columnwidth,clip]{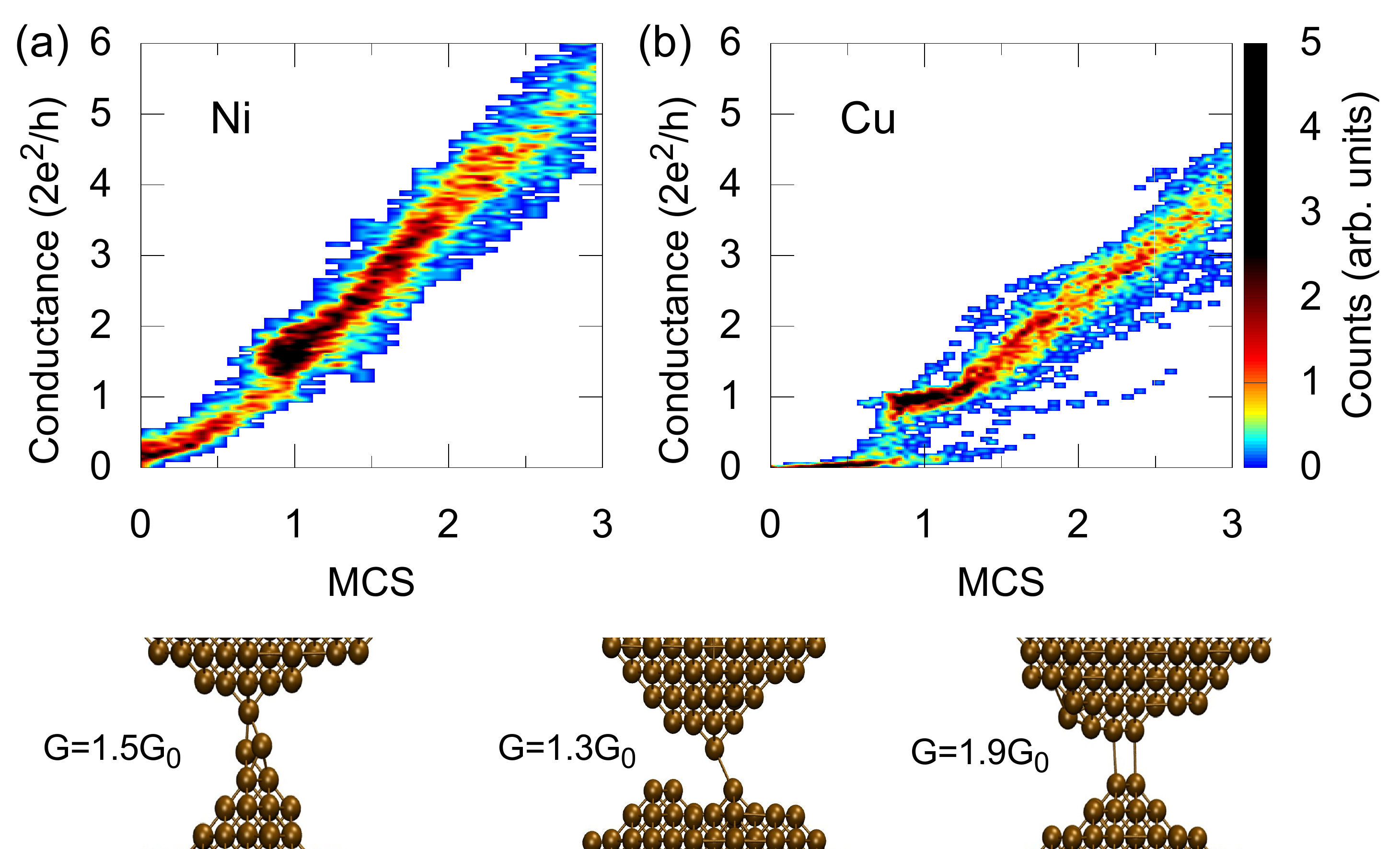} \end{center}
\caption{(Color online) Density plots of conductance vs.\ minimal cross section (MCS) for Ni (a) and Cu (b) obtained
from 100 simulated traces. Lower panels: Examples of simulated configurations of Ni atomic contacts that contribute
to the main peak in the conductance histogram, see Fig.~\ref{cond-histo}(e). The calculated conductance is indicated
next to each contact.}
\label{cond-vs-mcs}
\end{figure}

To understand the features observed in the conductance histograms, we turn to the analysis of the relation between conductance 
and structure in the simulated atomic contacts. An important structural parameter that is correlated with the transport 
properties is the minimal cross section (MCS) of the contact, whose precise definition was provided at the end of section 
\ref{sec-methods-thy}. Figures~\ref{cond-vs-mcs}(a,b) show 2D density plots of the MCS and conductance obtained from the 
contact geometries probed in our simulations for Ni and Cu. From these figures, we can see that the conductance of the 
last peak can be mainly ascribed to a contact with a single-atom constriction. For Ni, the conductance of the last peak 
($\sim 1.2$-$2.2G_0$) corresponds to a MCS of $0.7$-$1.3$, while for Cu the conductance range ($\sim 0.8$-$1.2G_0$) corresponds 
approximately to MCS values of $0.7$-$1.3$. We have obtained similar conclusions for Fe and Co contacts.

For all ferromagnetic contacts we find a large variance in the conductance for a given MCS value as 
compared with Cu. This observation indicates that the conductance of ferromagnetic contacts is more sensitive to the exact 
atomic configuration than in noble metals. Moreover, notice that for a given value of the MCS, the conductance of Ni contacts 
(and also Fe and Co contacts, not shown here) is higher than that of Cu contacts, which is clearly at variance with what 
happens in macroscopic wires of these materials.

The lower panels of Fig.~\ref{cond-vs-mcs} show three typical configurations of Ni contacts exhibiting conductance values 
in the range of the main conductance peak. We find two types of configurations which can be defined as single-atom
contacts: a single atom bonded to two or more atoms on each side (left contact) and contacts that show two under-coordinated 
atoms with a single bond on one side (central contact), which we will refer to as monomer and dimer, respectively. Note 
that configurations that cannot be defined as a single-atom contact (see right contact) also contribute to 
the conductance peak. In any case, the fact 
that these three different geometries contribute to the Ni conductance peak shows that it is not straightforward to 
establish a one-to-one correspondence between conductance peaks and atomic structures, especially in the case of 
multivalent metals. A similar variability is also found in the atomic configurations corresponding to the peaks in Fe and 
Co. In the case of Co, for which two peaks are observed, we find that the examined configurations contributing to the 
lower conductance peak can be mainly identified as single-atom contacts (\emph{i.e.}\ monomer and dimer configurations), 
while also other configurations can contribute to the higher conductance peak.

Interestingly, Figs.~\ref{cond-vs-mcs}(a,b) show a clear linear relation between the conductance and MCS, reflecting the 
dependence expected from Sharvin's conductance formula for a ballistic constriction \cite{Sharvin1965}. The observation 
that the linear relation extends almost all the way down to the single-atom level may seem quite surprising since, as 
we have seen above, the conductance is highly sensitive to small changes in the atomic geometry. The results of 
Figs.~\ref{cond-vs-mcs}(a,b) also show that there is a considerable variance of the conductance for a given MCS value. 
Thus for instance, the conductance for a Ni contact with a MCS of 1 atom can vary between $1G_0$ and $2G_0$. Therefore, 
the scaling of the MCS should be understood for the average conductance of many contact geometries.

\section{Shot noise} \label{sec-Noise}

\begin{figure}[b]
\begin{center} \includegraphics[width=\columnwidth,clip]{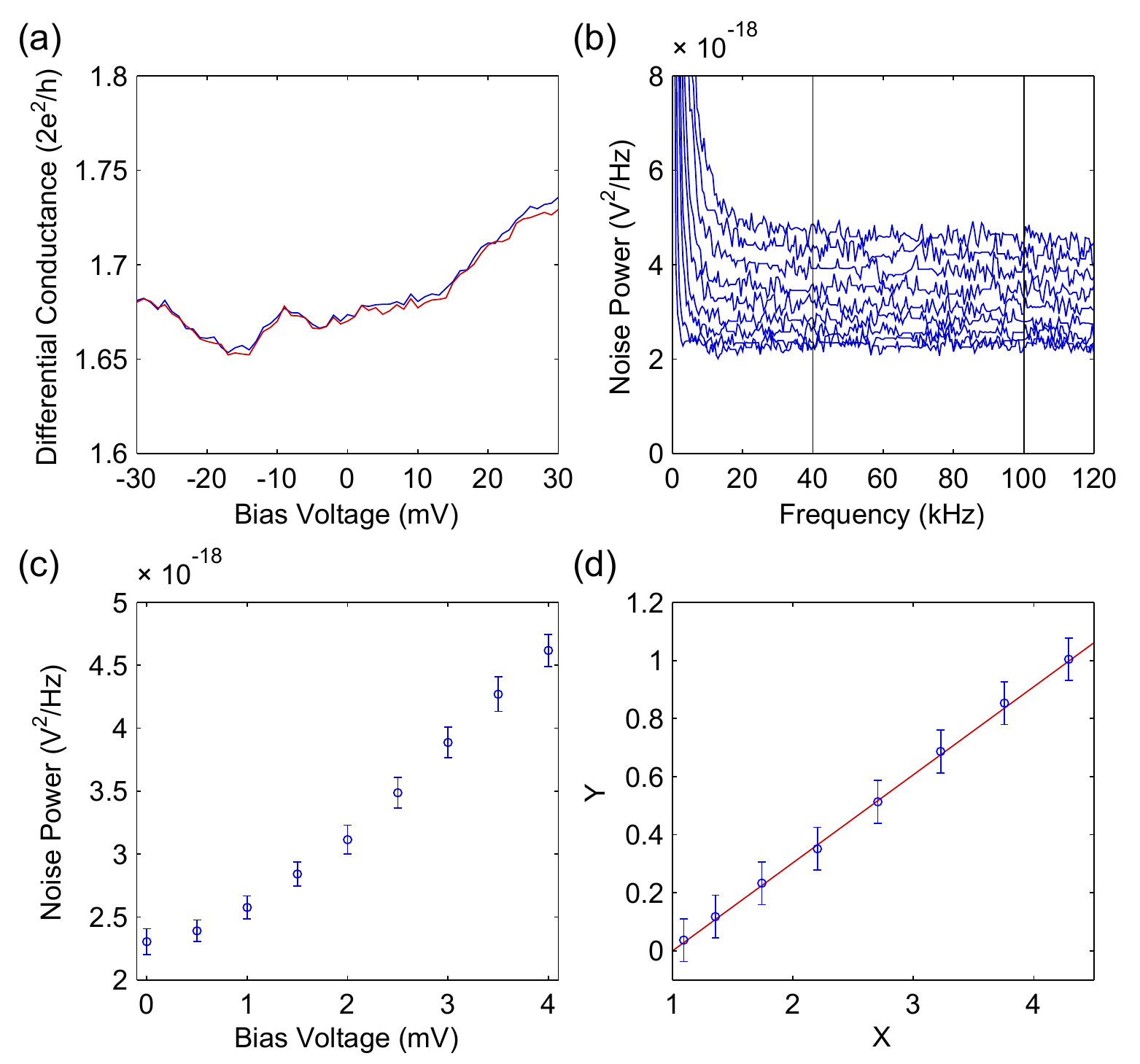} \end{center}
\caption{(Color online) (a) Differential conductance ($dI/dV$) curve measured for a Ni atomic contact before (blue) and 
after (red) shot noise measurements. (b) A series of noise spectra recorded at different bias voltages. Black lines 
indicate the frequency window which was selected for obtaining the average noise power. (c) Average noise power as 
function of bias voltage (d) Dependence of the reduced values $X,Y$ (blue) calculated for the measured noise in (c), 
and a linear fit (red), giving $F=0.30 \pm 0.01$ according to Eq.~(\ref{eq-Y}).}
\label{fig-noise}
\end{figure}
\begin{figure*}[t]
\begin{center} \includegraphics*[width=0.92\textwidth,clip]{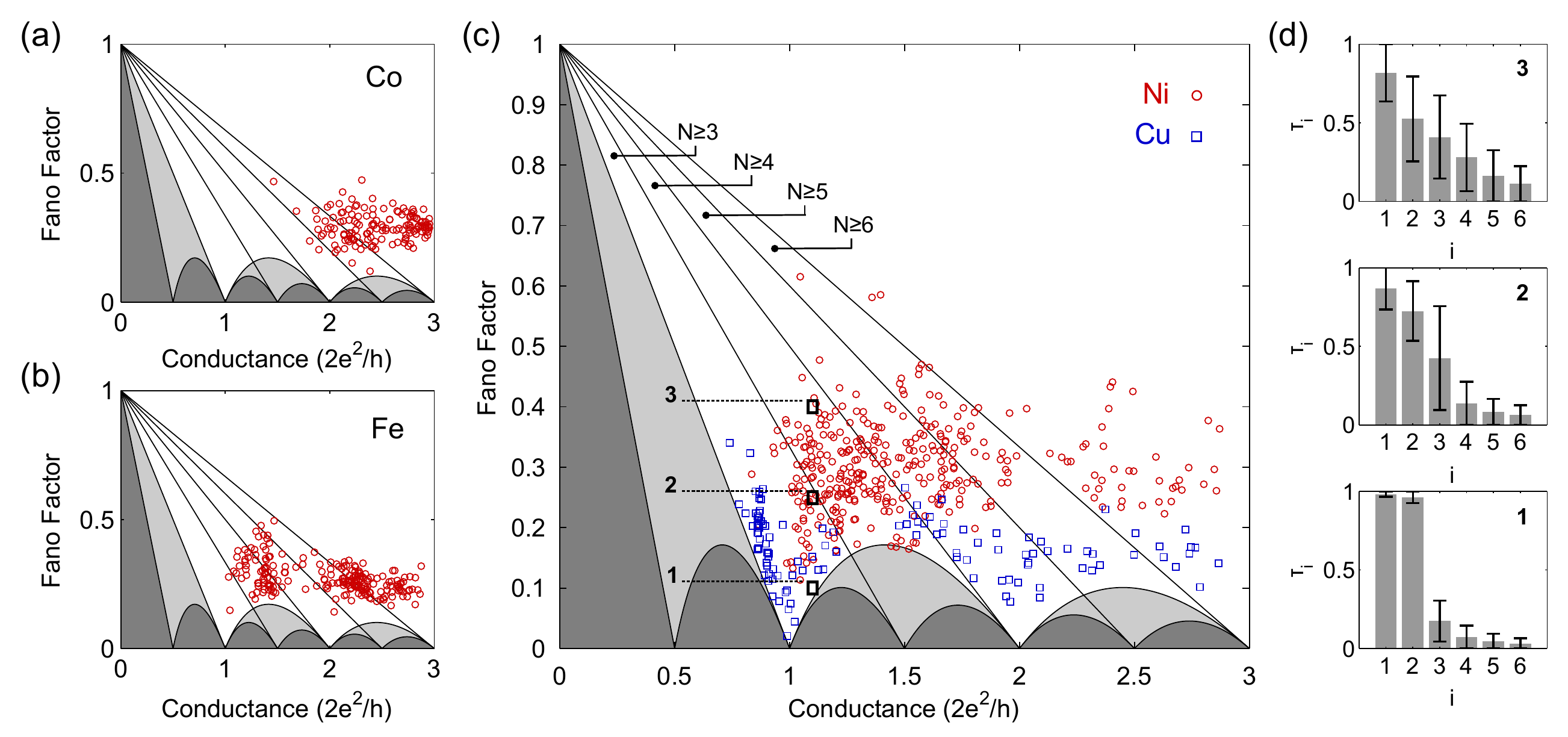} \end{center}
\caption{(Color online) (a-c) Fano factor as a function of the conductance as obtained from shot noise measurements 
for Co (a) Fe (b) Ni (c, red circles) and Cu (c, blue squares). Shaded areas show the inaccessible regions for the case 
of spin-degenerate transport (light shaded) and for the general case (dark shaded). Black lines starting from $(F,G)=(1,0)$
and ending at $(0,Ne^2/h)$ indicate the maximum $F$ obtainable for $N$ spin channels (see text). The measurement errors 
$\Delta G$, $\Delta F$ do not exceed $0.02G_0$ and $0.03$, respectively, for Cu and Ni, and $0.04G_0$ and $0.04$ for 
Fe and Co. (d) Spin-dependent conduction channel distributions for Ni contacts with measured values of $G=2.2 \pm 0.04e^2/h$, 
$F=0.40 \pm 0.01$ (top panel), $F=0.25 \pm 0.01$ (middle), $F = 0.10 \pm 0.01$ (bottom). These values of conductance 
and Fano factor are indicated as black squares in (c). Error bars indicate minimum and maximum possible transmission 
for each spin channel.}
\label{fig-Fano}
\end{figure*}

To further investigate the transport properties of ferromagnetic contacts, we have conducted measurements of the electronic 
shot noise generated by the contacts. Let us remind that within the Landauer-B\"uttiker framework, shot noise depends on 
the number of open conduction channels and their transmission probabilities, see section IIB. The overall zero-frequency 
current noise generated by a quantum conductor (including the thermal noise) can be expressed as \cite{Blanter2000,Kumar2012}
\begin{equation}
\label{eq-SI}
S_I = 4k_{\rm B} TG \left[ 1 + F (x \coth(x) - 1) \right] ,
\end{equation}
where $G$ and $F$ are the conductance and Fano factor, respectively [see Eqs.~(\ref{eq-G-S}) and
(\ref{eq-Fano})], and $x=eV/2k_{\rm B} T$ describes the ratio between bias voltage $V$ and temperature $T$.
Near equilibrium conditions ($x \ll 1$), $S_I$ reduces to the Johnson-Nyquist expression for thermal noise
$4k_{\rm B} TG$. At high bias ($x \gg 1$), the current noise depends linearly on the bias current, $S_I=2eIF$.
Figure~\ref{fig-noise} shows an example shot noise measurement for a Ni atomic contact. Differential conductance
spectra ($dI/dV$) of the contact, see Fig.~\ref{fig-noise}(a), are recorded before and after the noise measurement 
in order to confirm that the junction has remained stable during the measurement. The zero bias conductance of the
junction is determined from the average differential conductance in the window of $|V|<5$ mV. Figure~\ref{fig-noise}(b)
shows a series of noise spectra for different applied bias. Each spectrum is obtained from the Fourier transform
of voltage fluctuations produced by the junction, and averaged for 5000 consecutive measurements. The voltage
noise produced by the amplifier was measured separately and subtracted from the recorded spectra. The spectra
were corrected to account for low pass filtering due to capacitance of the cabling and amplifier input capacitance
(total capacitance of $\sim 40$ pF). The noise power is averaged in a frequency window, which is selected to be high 
enough to reduce $1/f$ noise contributions. Figure~\ref{fig-noise}(c) shows the average noise power as function of 
bias voltage across the junction. Following Ref.~[\onlinecite{Kumar2012}], Eq.~(\ref{eq-SI}) can be expressed as
\begin{equation}
\label{eq-Y}
Y = F (X-1) ,
\end{equation}
where $X=x \coth(x)$ and $Y=[S_I (V)-S_I (0)] / S_I (0)$. The Fano factor is obtained by calculating the
reduced parameters $X,Y$ and obtaining a linear fit of $Y(X)$ according to Eq.~(\ref{eq-Y}), as shown in
Figure~\ref{fig-noise}(d).

Figure~\ref{fig-Fano} shows the distribution of $F$ as function of zero-bias conductance $G$ obtained for Fe,
Co, Ni, and Cu atomic contacts. For Cu, a suppression of the Fano factor towards its lower limit for spin-degenerate 
transmission (light shaded area) is observed. This behavior is similar to the results obtained
for the monovalent metals Au and Ag, indicating conductance quantization, \emph{i.e.}, that in multiples of $G_0$
the conductance takes place through nearly fully-open channels \cite{vandenBrom1999,Vardimon2013}. Conversely, a distinct
qualitative picture arises for ferromagnetic contacts. The measured values of $F$ are found to be significantly higher
compared to Cu. In particular, the large suppression at multiples of $G_0$ observed for Cu is clearly missing for the 
ferromagnetic contacts.

The conductance and Fano factor depend on the distribution of conduction channels carrying the current, in accordance 
with Eqs.~(\ref{eq-G-S}) and (\ref{eq-Fano}). This dependency allows us to draw several conclusions regarding the set 
of transmission probabilities $\{ \tau_{n\sigma} \}$. First, one can determine the minimum number of channels contributing 
to transport according to the position of the measurement in the $(F,G)$ space. The solid lines starting from $(F,G) = (1,0)$ 
and ending at $(0,Ne^2/h)$ in Fig.~\ref{fig-Fano}(a-c) indicate the maximum Fano factor that can be obtained
for $N$ spin channels (\emph{i.e.}\ having a maximum conductance of $Ne^2/h$). Therefore, given a combination
of $(F,G)$ one can obtain a lower bound for the number of transmitting channels (indicated within the regions between
each two lines in Fig.~\ref{fig-Fano}(c)). For Ni, one can see that a minimum of $N=3$-$6$ spin channels contribute
to transport for contacts with conductance values corresponding to configurations with single-atom cross section
(\emph{i.e.}\ $\sim 1.2$-$2.2G_0$, see Fig.~\ref{cond-histo}(e)). Similarly, the results for Co and Fe indicate that 
the current through a single-atom contact is carried through a multiple number of channels.

To study the transmission values $\{ \tau_{n\sigma} \}$, we use a numerical analysis introduced in Ref.~[\onlinecite{Vardimon2013}].
The analysis is based on enumerating the possible combinations of $\{ \tau_{n\sigma} \}$ up to a certain number of 
channels $N$ (here, $N=6$ was chosen to account for the main channels), and
identifying the range of solutions that give the measured $F$ and $G$. In many cases, this analysis allows us to determine
the transmission probabilities with reasonable accuracy. We note that since Eqs.~(\ref{eq-G-S}) and (\ref{eq-Fano}) are
symmetric with respect to spin, the spin direction cannot be determined using this method. The transmission probabilities
are labeled by a single index $i$, and are ordered according to decreasing transmission. Figure~\ref{fig-Fano}(d) shows the possible
values of $\{ \tau_{n\sigma} \}$ for three different combinations of $F$ and $G$ for Ni contacts, which are indicated in
Fig.~\ref{fig-Fano}(c). For combination 1 (bottom panel), in which the Fano factor is near its minimum limit, two fully-open
spin channels carry most of the current. In the case of Cu, for which spin polarization is not expected, measurements located
at this position indicate transport dominated by two channels (one for each spin type) with identical transmission. As can be 
seen from examples 2 and 3 (middle and upper panels in Fig.~\ref{fig-Fano}(d)), an increase in $F$ results in a larger number 
of partly-open channels carrying the current. This trend appears for any conductance range; a low Fano factor
near the minimum limit indicates a channel distribution with a minimal number of channels, and maximal number of open
channels ($\tau_{n\sigma}=1$), while a higher $F$ results in a distribution with a larger number of channels and lower
transmission values. Thus, the large spread in $F$ observed for Fe, Co, and Ni implies that there is no conductance quantization,
while the number of channels and their transmission probabilities vary significantly between different contact realizations.

Our analysis of conductance traces and shot noise measurements indicates that the conductance and channel distribution 
in ferromagnetic contacts are highly sensitive to atomic geometry. This sensitivity makes the comparison between theory and 
experiment challenging. Interestingly, the structure-conductance analysis obtained from our simulations (Fig.~\ref{cond-vs-mcs}), 
shows that the relative variance of the conductance in respect to the mean value ($\Delta G / G$) is greatly reduced 
with increasing MCS. Thus, the study of larger contacts could be advantageous as the sensitivity to atomic geometry is 
significantly reduced. Unfortunately, the conductance histograms for ferromagnetic contacts do not show 
clear features at values above $\sim 4G_0$. On the other hand, the study of shot noise at larger conductance values could 
potentially provide new information, allowing a more straightforward comparison with theory.

\begin{figure}[t]
\begin{center} \includegraphics[width=0.95\columnwidth,clip]{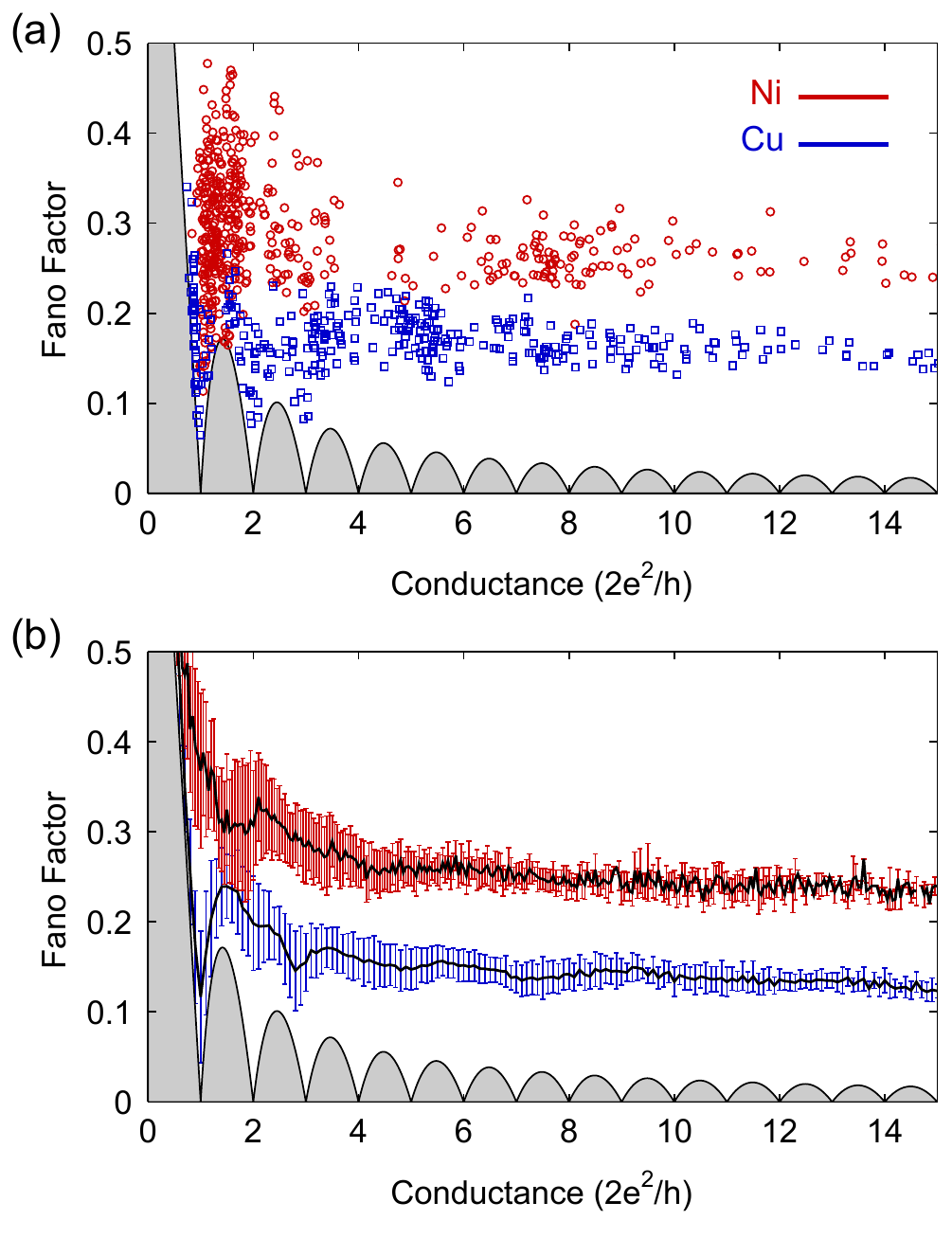} \end{center}
\caption{(Color online) Saturation of Fano Factor. (a) Measured Fano factor as a function of the conductance for Ni 
contacts (red) and Cu contacts (blue). (b) The corresponding computed Fano factor as a function of conductance. The solid
lines indicate the average value, while error bars show the corresponding standard deviation.}
\label{fig-Fano-sat}
\end{figure}

To investigate this possibility, we have extended our measurements and simulations to contacts with higher conductance values,
focusing on Ni and Cu for comparison between a ferromagnetic metal and a monovalent metal. Figure~\ref{fig-Fano-sat}(a) shows
the distribution of the Fano factor as function of conductance measured up to $15G_0$, for Ni and Cu contacts. Remarkably,
for both metals, the variance in the Fano factor is strongly reduced for larger contacts, and the Fano factor saturates to a 
nearly constant value of $0.26 \pm 0.02$ and $0.15 \pm 0.01$ for Ni and Cu, respectively.

\begin{figure*}[t]
\begin{center} \includegraphics*[width=\textwidth,clip]{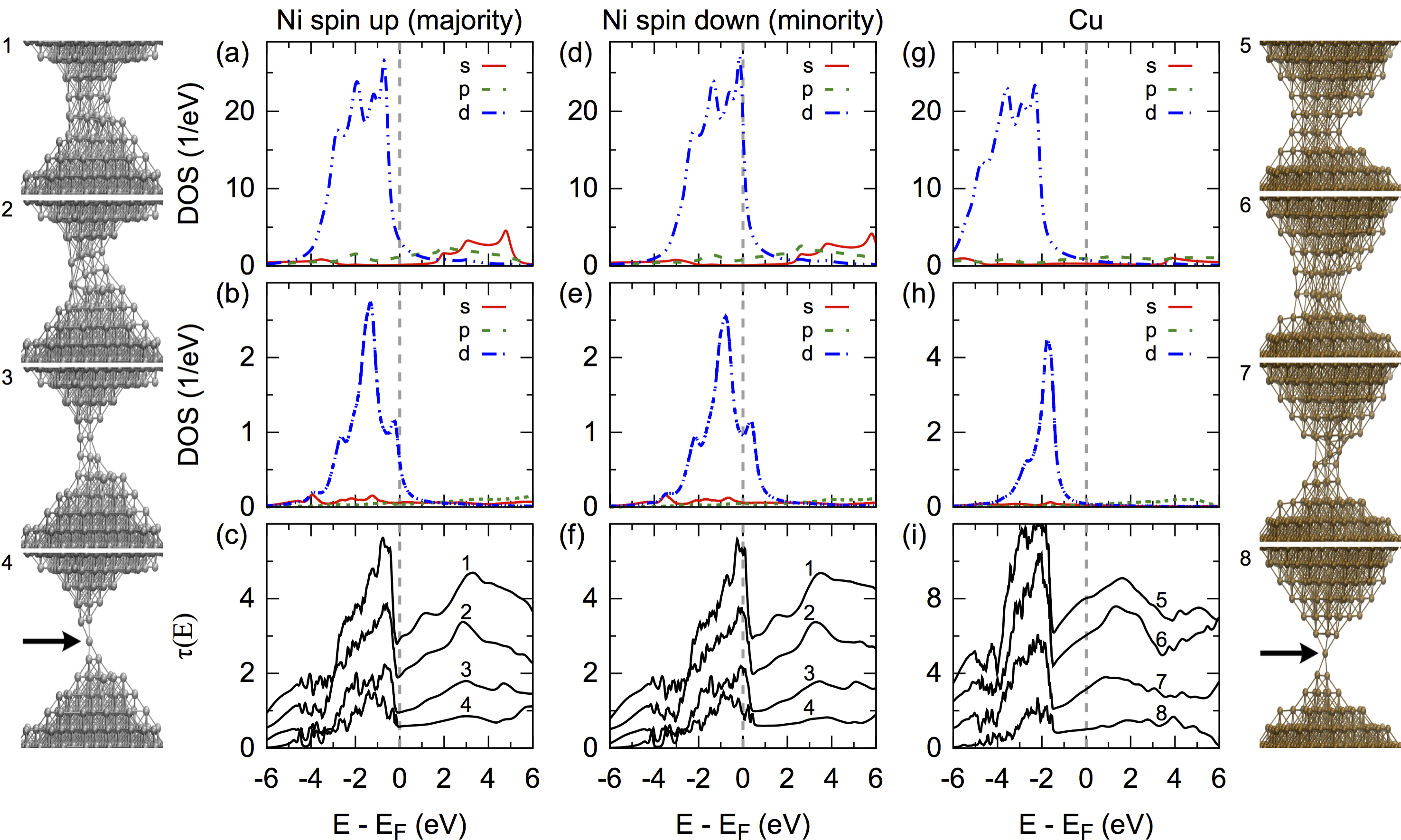} \end{center}
\caption{(Color online) (a) Bulk density of states (DOS) as a function of energy for the majority-spin (or spin-up)
bands of Ni. We show the total contribution of the five $3d$ orbitals, the $4s$ orbital, and the three $4p$
orbitals. (b) The energy dependence of the local DOS for majority spins projected onto the central atom (see arrow)
of the Ni single-atom contact number 4 shown in the left column. Again, the different curves correspond to the total
contributions of the $s$, $p$, and $d$ orbitals. (c) The total transmission as a function of energy for the
majority spins for the four Ni contacts shown in the left column which were obtained in one of the simulations
of the breaking process. (d-f) The same as in panels (a-c) for the minority-spin electrons. (g-i) The same as
in panels (a-c) for Cu. The corresponding Cu contact geometries are shown in the right column. In all panels, the
vertical dashed lines indicate the position of the Fermi energy, as a guide to the eyes.}
\label{fig-DOS}
\end{figure*}

Figure~\ref{fig-Fano-sat}(b) shows the $(F,G)$ distribution obtained from our simulations. The results are found to be in 
very good agreement with the experimental values, both in the large variance at low conductance, and in the saturation at 
higher conductance values. The simulated saturation values obtained yield $0.23 \pm 0.01$ and $0.13 \pm 0.01$ for Ni and Cu, 
respectively. Our simulations for Fe and Co result in similar behavior to that obtained for Ni, both giving the same saturation 
value of $0.27 \pm 0.01$, see Appendix A. For Au, having a monovalent orbital structure similar to Cu, the Fano factor 
obtained by our simulation reaches a value of $0.15 \pm 0.02$. We note that previous measurements for Au contacts have 
indicated an average $F$ of $\sim 0.15$ for $G>10G_0$ \cite{vandenBrom2000}. Thus, the clearly larger Fano factor values 
encountered in the ferromagnetic metals as compared to the noble ones seem to be a generic feature, which reflects the 
different orbital structure of these two kinds of metals. This observation will be further discussed in the following section.

\section{Nature of the conduction channels} \label{sec-Channels}

The goal of this section is to elucidate the transport mechanism in ferromagnetic atomic contacts by analyzing the electronic
structure and conduction channels of our simulated atomic contacts. We shall show that the fundamental differences between
noble and ferromagnetic metals are related to the significant role that $d$ orbitals play in the transport properties in
the latter case. Furthermore, we find that the exchange-splitting of the $d$ orbitals results in distinct transport
for majority-spin and minority-spin electrons. For didactic reasons, in this section we focus on the comparison between Ni
and Cu. We find that the conclusions drawn for Ni are qualitatively valid for Fe and Co. Our theoretical results for these
metals are shown in Appendix A.

Figure~\ref{fig-DOS} shows a comparison of the energy dependence of both the density of states (DOS) and the transmission 
function for Ni and Cu, as calculated from our tight-binding model. The upper panels show the bulk DOS, while the middle 
panels display the local DOS projected onto the central atom for a single-atom contact configuration. The contributions 
of $3d$, $4s$ and $4p$ orbitals to the DOS are shown in separate curves. In the case of Ni, we show separately the 
contributions of the majority-spin (or spin-up) electrons and minority-spin (or spin-down) electrons to both DOS and
transmission. Comparing these figures, one can immediately see that the main difference between Ni and Cu is the large 
contribution of $d$ orbitals at the Fermi energy for Ni. This contribution is particularly large in the case of 
minority-spin electrons, for which the $d$ states are located higher in energy due to the spin-splitting for the bulk $d$ bands.

The lower panels of Fig.~\ref{fig-DOS} show the total transmission as a function of energy for four different contacts
with different cross sections obtained in an individual simulation of the breaking process of a Ni and a Cu wire. The
corresponding contact geometries are shown in the left (Ni) and right (Cu) columns. As one can see in Fig.~\ref{fig-DOS}(c,f),
the corresponding spin-resolved total transmission follows closely the energy dependence of the total DOS. Thus for instance,
the total transmission at the Fermi energy, $\tau(E_{\rm F})$, for the minority-spin electrons is significantly higher than
that of the majority-spin electrons, in correlation with the higher DOS at $E_{\rm F}$ in the former case. These results
strongly suggest that $d$ orbitals are responsible for the significantly higher transmission for minority-spin electrons.
Importantly, we find that the key role of the $d$ orbitals in transport properties also extends to atomic contacts with
larger contact sizes.

\begin{figure}[t]
\begin{center} \includegraphics[width=\columnwidth,clip]{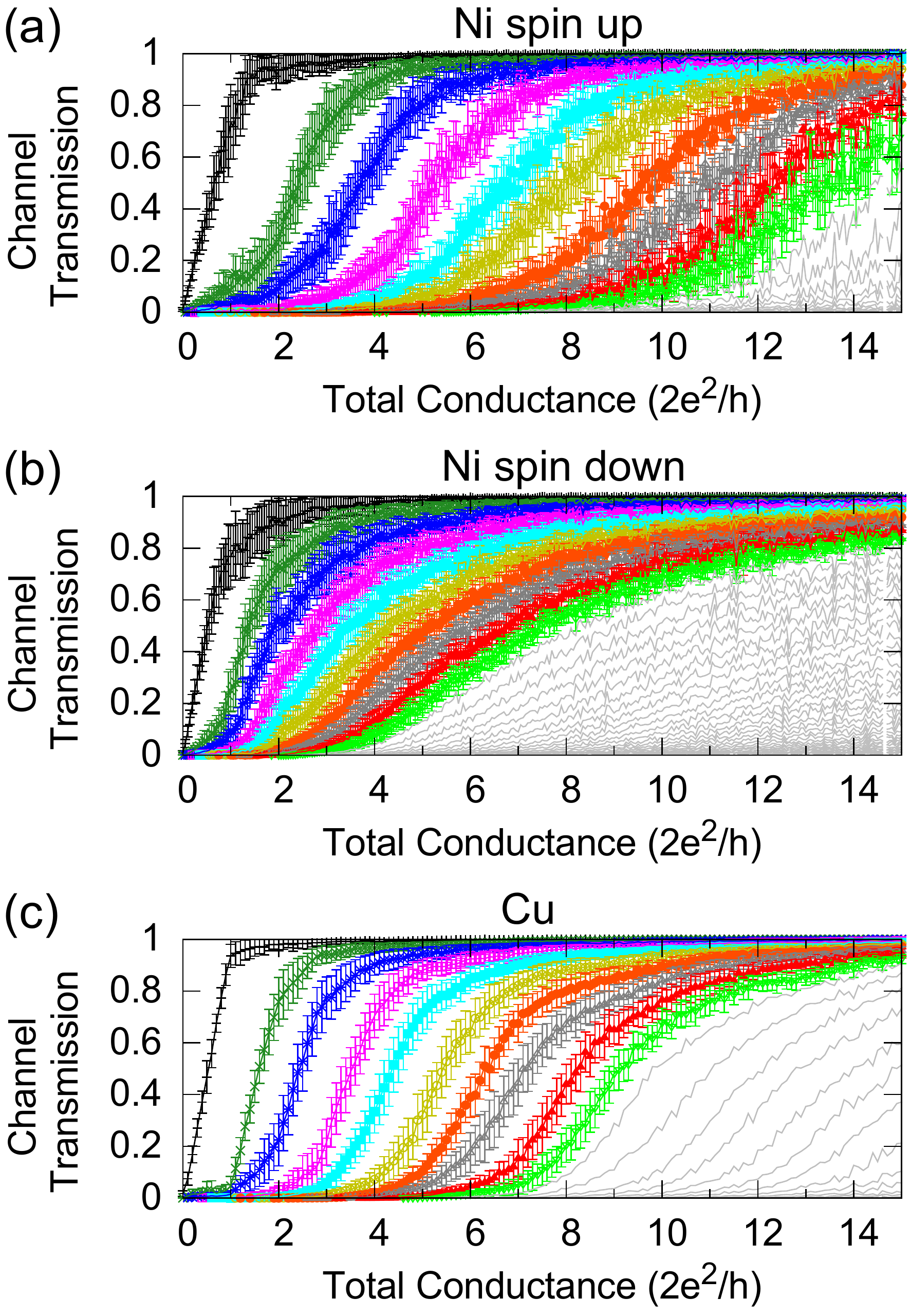} \end{center}
\caption{(Color online) (a) Transmission coefficients for Ni majority-spin (spin-up) electrons as a function
of the contact conductance. The lines inicate to the average values and the bars to the standard deviations.
(b) The same as in panel (a) but for Ni minority-spin (spin-down) electrons. (c) The corresponding results for Cu.}
\label{fig-channels}
\end{figure}

To further study the transport mechanism in ferromagnetic contacts, we examine the distribution of conduction channels.
Figure~\ref{fig-channels} shows the average value of the transmission probabilities as a function of conductance as
obtained from 100 contact stretching simulations for Ni (spin up and down) and Cu. In the case of Ni, one can immediately
see that there is a larger number of channels contributing to transport for minority-spin electrons. For example, for
conductance values related to a single-atom Ni contact (\emph{i.e.}\ $\sim 1.2$-$2.2G_0$), there are approximately
4-7 minority-spin channels compared to 2-4 majority-spin channels. The larger number of minority-spin channels can be
correlated with the higher number of the minority-spin $d$ states at $E_{\rm F}$ (Fig.~\ref{fig-DOS}(d,e)), which are
available for conduction electrons.

In the case of Cu (Fig.~\ref{fig-channels}(c)), we see that transport is dominated by channels with nearly perfect
transparency which open one by one with increasing conductance. This behavior naturally explains the suppression of
shot noise at multiples of $G_0$ observed for this metal. As the conductance increases (or equivalently the contact size)
more partially-open channels appear, explaining why the Fano factor does not vanish for contacts with conductance above
$3G_0$. Interestingly, a tendency for transport through nearly-open channels is also observed for Ni spin-up electrons
(Fig.~\ref{fig-channels}(a)). However, as the transport is mainly carried by spin-down channels with intermediate
transmission values, one does not observe any strong suppression of the Fano factor at any particular value of the
conductance.

\begin{figure}[t]
\begin{center} \includegraphics[width=0.9\columnwidth,clip]{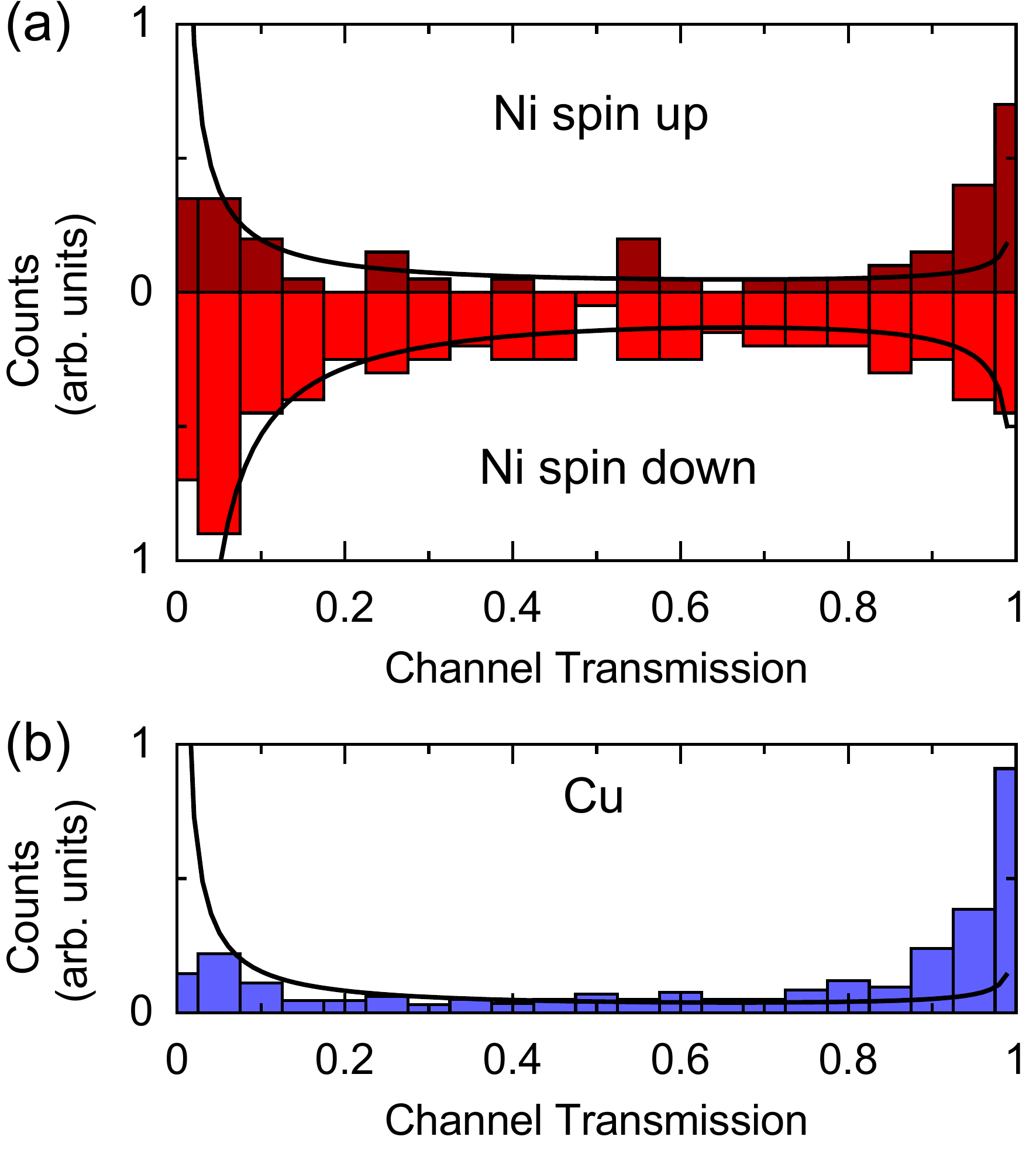} \end{center}
\caption{(Color online) (a) Transmission coefficient histogram for Ni contacts with a conductance of $15G_0$. We
show the results for both majority-spin (spin-up) electrons and minority-spin (spin-down) electrons. (b) The
corresponding results for a Cu with $G=15G_0$. In both panels the dotted line shows the bimodal distribution 
of Eq.~(\ref{eq-bimodal}).}
\label{fig-distr}
\end{figure}

To understand the origin of the saturation of the Fano factor and the differences in the saturation values between
Cu and the ferromagnetic metals, we show in Fig.~\ref{fig-distr} histograms of the transmission eigenvalues for Ni
and Cu contacts with a conductance of $15G_0$. As one can see in Fig.~\ref{fig-distr}(a), the spin-up and spin-down
transmission coefficients are distributed very differently. In the former case, the distribution shows a preference 
to open channels. In contrast, for minority spins, the distribution is mainly composed by partially open channels. As
discussed in section \ref{sec-Noise}, the Fano factor increases as the conductance is determined by a larger number of
partially-open channels. Thus, from the transmission histograms, we can conclude that the lower value of the Fano 
factor at saturation in the case of Cu results from the dominant role of nearly fully open channels, while for Ni this 
value is significantly larger due to the presence of numerous partially open (minority-spin) channels.

The results presented so far indicate that transport for Ni spin-up and spin-down electrons differs not only in the total
transmission value, but exhibits fundamentally distinct behaviors. The transmission distribution for the majority spins
clearly resembles that of the Cu contacts (Fig.~\ref{fig-distr}(b)), in which transport is dominated by highly transmissive
channels. This behavior can be traced back to the fact that in both cases (Cu and Ni spin up), the states available for
transport originate mainly from the $s$ valence orbitals. On the other hand, the channel distribution of the Ni minority
spins is very similar to that of the transition metal Pt \cite{Evangeli2015}, with dominant $d$ valence orbitals near the 
Fermi energy. Thus, we see that Ni behaves in some sense as a combination of monovalent and transition metal conductors 
in parallel.

As it is well known in mesoscopic physics, the Fano factor of a metallic diffusive wire is universal (it does not depend
on the degree of disorder or on the wire geometry) and adopts a value equal to $1/3$ \cite{Nagaev1992,Beenakker1992}.
This value can be understood in terms of the distribution of transmission coefficients, which in the diffusive wire case
reduces to the so-called bimodal distribution \cite{Dorokhov1982,Beenakker1997}
\begin{equation}
\label{eq-bimodal}
P(\tau) = \frac{G}{G_0} \frac{1}{2 \tau \sqrt{1 - \tau}} ,
\end{equation}
where $P(\tau)$ is the probability density to find a given value of the transmission $\tau$ and $G$ is the wire conductance.
To understand the differences between our atomic contacts and the diffusive wire case, we show in the histograms of
Fig.~\ref{fig-distr} the bimodal distribution of Eq.~(\ref{eq-bimodal}) as a solid line. As one can see in
Fig.~\ref{fig-distr}(a), the distribution of the Ni minority spins follows the bimodal distribution. Indeed, the Fano factor
for the minority-spin electrons $F_{\downarrow} = \sum{\tau_{n,\downarrow}(1-\tau_{n,\downarrow})}/\sum{\tau_{n,\downarrow}}$ 
yields 0.28, 0.30 and 0.34 for Ni, Co and Fe, respectively. These values are very close to the universal value of 1/3. 
Thus, our results suggest that ferromagnetic atomic contacts show a
diffusive-like behavior for minority-spin electrons. On the other hand, the transmission distributions for the Ni majority
spins and for Cu differ markedly from the bimodal distribution, the main difference being the presence of a larger number
of channels with very high transmissions. This fact leads to a reduction of the Fano factor as compared to the diffusive case.
In this respect, we find that the Fano factor for the Ni majority-spin electrons is $F_{\uparrow} = 0.16$, while
$F_{\uparrow} = 0.22$ for Co, and $F_{\uparrow} = 0.20$ for Fe. These values are closer to the saturation value for Cu,
$F=0.15$. We will further discuss these results in the following section.

\begin{figure}[t]
\begin{center} \includegraphics[width=0.9\columnwidth,clip]{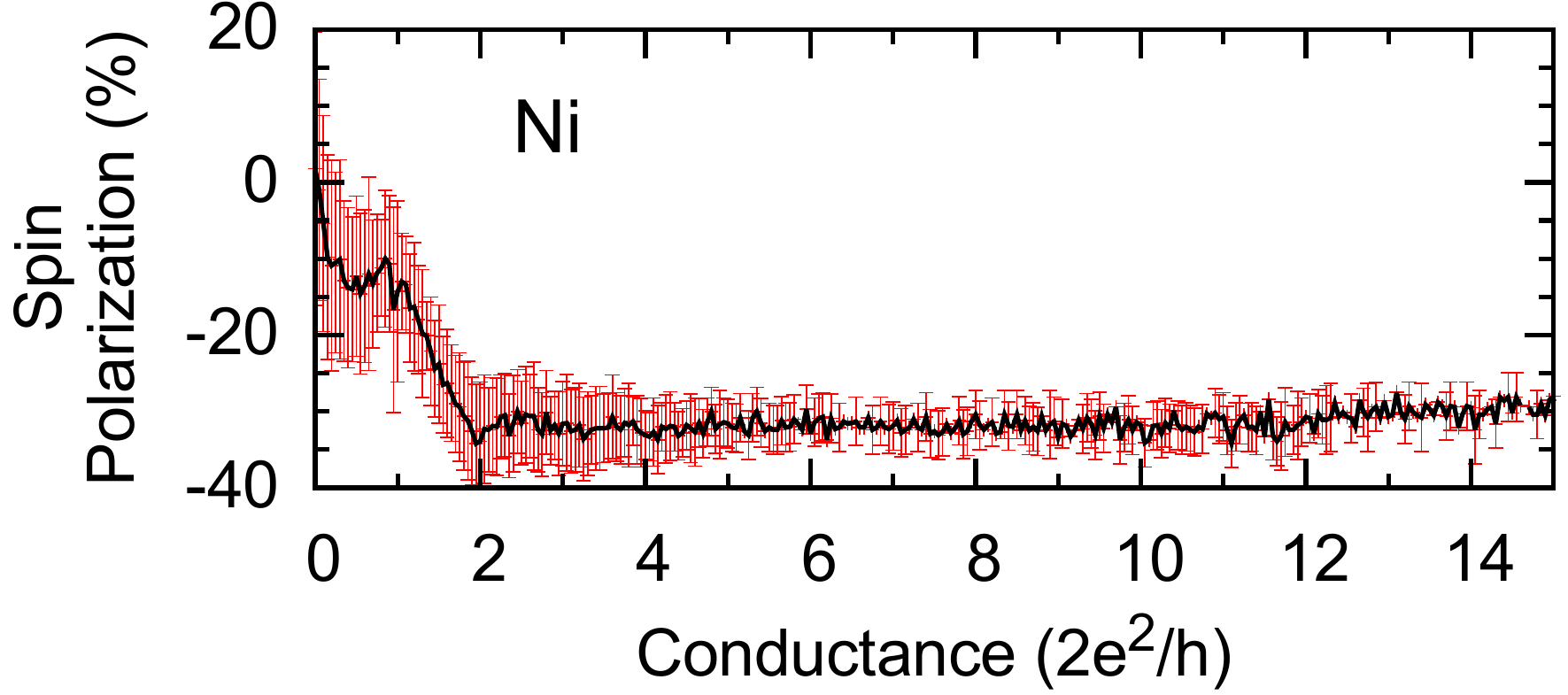} \end{center}
\caption{(Color online) Computed spin polarization of the current as a function of the conductance for
Ni. The solid lines correspond to the average value and the bars to the standard deviation.}
\label{fig-spinpol}
\end{figure}

Another aspect of the conduction across ferromagnetic atomic contacts that differs dramatically from the macroscopic
case is the spin polarization of the current. This quantity is defined as
\begin{equation}
P_I = \frac{I_{\uparrow} - I_{\downarrow}}{I_{\uparrow} + I_{\downarrow}} \times 100\% ,
\end{equation}
where $I_{\uparrow}$ is the current carried by the majority-spin electrons and $I_{\downarrow}$ is the current carried
by the minority-spin electrons. In the case of macroscopic wires, Mott's model predicts a positive sign for the spin
polarization because transport is dominated by majority-spin electrons, as explained in the introduction. The positive spin
polarization was found to be consistent with studies of sub-gap structure measurements of ferromagnet-superconductor interfaces
\cite{Meservey1994}. In contrast, following our discussion above, it is clear that in the case of atomic-size contacts, the
current is dominated by minority-spin electrons.The influence of this property is illustrated in Fig.~\ref{fig-spinpol}, 
showing the current spin polarization as a function of the conductance. As one can see, the spin polarization is negative, 
irrespective of the conductance value, and it saturates to a value of $\sim -30\%$ for large contacts. Interestingly, 
while the presence of spin-down $d$ states at the Fermi energy reduces the minority-spin conductivity in the bulk case, 
for atomic contacts the situation is precisely the opposite: additional minority-spin channels with $d$ character become 
available for transport, resulting in an opposite sign of the spin polarization.

\section{Discussion} \label{sec-further-discussions}

Our results from both experiment and theory clearly indicate that $d$ orbitals are responsible for fundamental differences
in transport properties of ferromagnetic contacts compared to monovalent contacts. Ferromagnetic contacts clearly differ 
from Cu (and other monovalent metals) by: (i) higher conductance, (ii) larger variance in the conductance and shot noise 
(for small contacts), (iii) existence of multiple partly-transmitting channels, and (iv) nearly double Fano factor saturation 
value for large contacts. As our channel analysis reveals, transport in ferromagnetic contacts can be viewed to occur in 
parallel through spin-up channels, behaving similar to a monovalent metal, and spin-down channels, acting as a transition 
metal. Thus, the aforementioned properties can be mainly attributed to the contribution of minority-spin electrons, for 
which $d$ states are available for transport.

We will therefore divide our discussion into $s$-dominated transport (\textit{e.g.}\ Cu, Ni spin up) and transport through 
mixed $sd$ states (\textit{e.g.}\ Ni spin down). In the case of $s$-dominated transport, conductance quantization, 
\textit{i.e.}, transport through fully-open channels, takes place only for small contacts ($G\lesssim 3G_0$), whereas for 
larger contacts, an increasing contribution of partially-open channels is observed. This behavior was reasonably 
reproduced by a model of a free-electron constriction with disorder, connecting two gradually narrowing contacts 
\cite{Burki1999,Burki1999b}. The mentioned disorder can be associated with local impurities, scattering at the contact surfaces 
or lack of periodical lattice structure. Indeed, one could naturally expect that as the constriction size increases, the 
series resistance induced by disorder will play a more dominant role in transport, obscuring the measurement of 
quantized values of conductance. We argue that the saturation of the Fano factor could be viewed as an interplay between 
the quantization of the contact on one hand and some amount of disorder which results in imperfect transmissions. These 
arguments explain why the average value of the Fano factor ($\sim$ 0.15) is significantly lower than the value of 1/3 
expected for a diffusive contact.

In the case of transport with $sd$-character, our results suggest that disorder is significantly higher compared to 
a monovalent conductor. We find this expressed in the bimodal distribution observed for spin-down Ni channels (and the
corresponding $F_{\downarrow} \sim 1/3$), being a characteristic of diffusive transport. A higher disorder in 
ferromagnetic contacts can be related to the inherent anisotropy of $d$ orbitals compared to the high-symmetry of $s$ 
orbitals. The interplay between the disorder in diffusive wires and a ballistic quantum point contact has been studied 
theoretically by Beennakker and Melsen \cite{Beenakker1994}. The model analyzed in that work was able to recover the 
saturation of the Fano factor to 1/3 with increasing contact size, however the channel distribution obtained for 
narrow contacts is at variance with the partially-open channel distribution we obtain. This discrepancy indicates 
the need to consider the specific orbital structure of the contacts.

We note that Riquelme \emph{et al.}\ \cite{Riquelme2005} reported an experimental analysis of the transmission 
distribution of Pb atomic contacts with conductance values ranging between $1G_0$ and $15G_0$. The study was 
based on the use of superconductivity and the analysis of the sub-gap current, in the spirit of 
Refs.~[\onlinecite{Scheer1998,Scheer1997}]. The authors concluded that as the contact size increased, the 
transmission distribution approached very quickly the universal bimodal distribution. Those authors also presented 
a theoretical analysis based on ideal geometries and a tight-binding model similar to the one employed here that 
suggested that this behavior could be associated with the $sp$ valence orbital structure of Pb that gives rise to 
a significant contribution of partially-open channels even in the absence of atomic disorder. This conclusion, 
together with our results, suggests that such behavior may be a general characteristic of multivalent metals.

When discussing diffusive transport in atomic contacts, it is important to consider the conductor dimensions. 
The strict definition of diffusive transport requires that the length of the conductor satisfies $L \gg \ell$, 
where $\ell$ is the elastic mean free path. In the case of atomic contacts made of the monovalent metal Au, 
different estimates of $\ell$ give a value of approximately 4-5 nm \cite{Ludoph1999,Ludoph2000c,Erts2000}. 
In comparison, the length of the contacts in our simulations is up to 3 nm in total. Since the saturation of 
the Fano factor is captured in our simulated structures, we infer that this region is sufficient to describe 
the main transport characteristics. Thus, it is rather unlikely that the condition $L \gg \ell$ for diffusive 
transport is met. Yet, we can find two reasons why diffusive behavior is seen nevertheless. First, the increased 
disorder originating from the contribution of $d$ orbitals could lead to a considerably lower value for $\ell$ 
compared to Au. Second, theoretical studies have shown that it is sufficient to have a small number of tunneling 
barriers in series to result in Fano factor near 1/3 \cite{deJong1995}. These considerations can explain why 
the bimodal channel distribution can appear even in contacts of several atoms in cross section.

\begin{figure*}[t]
\begin{center} \includegraphics*[width=0.9\textwidth,clip]{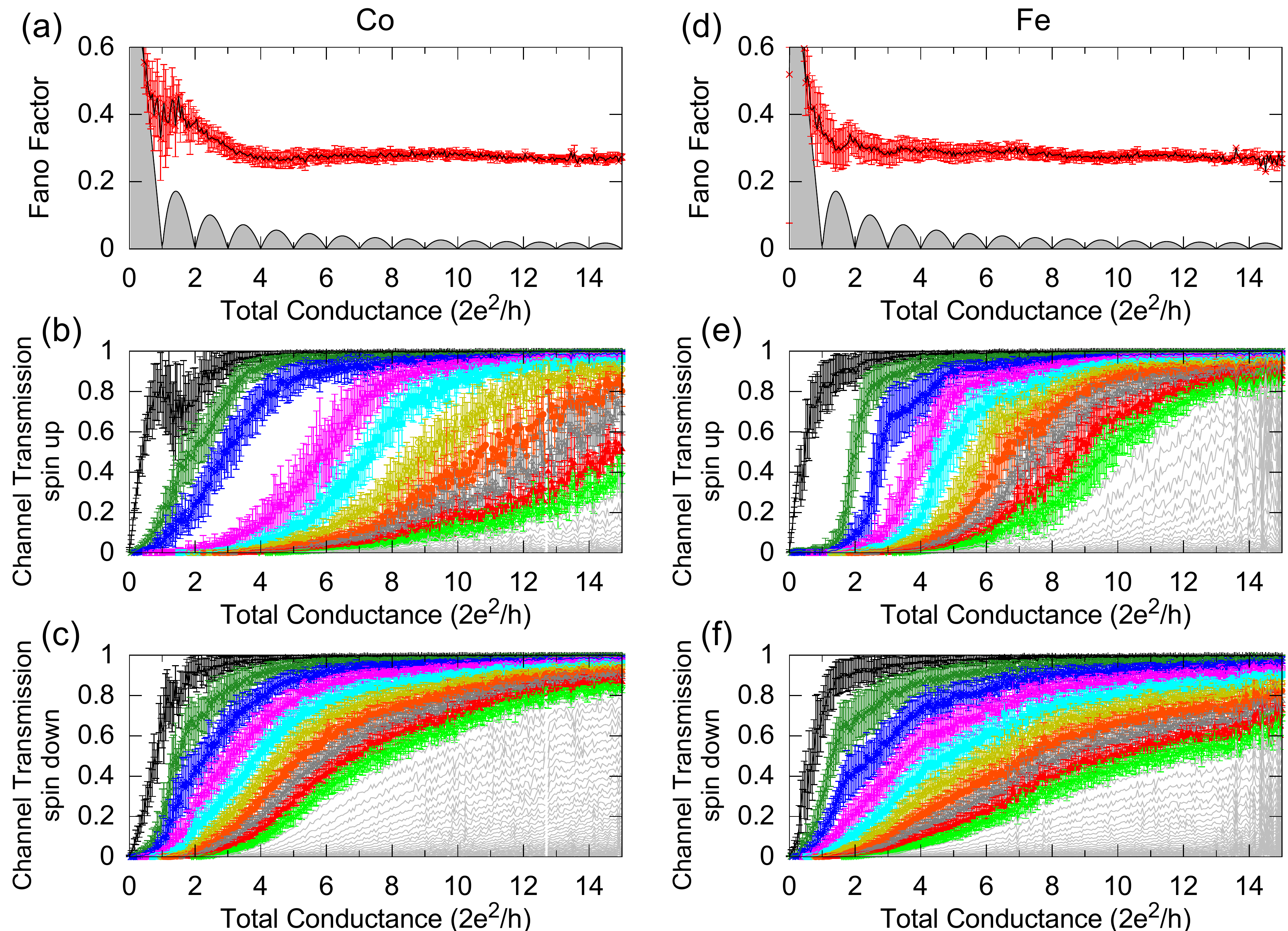} \end{center}
\caption{(Color online) (a) Computed Fano factor as a function of the conductance for Co contacts. The
solid (black) line corresponds to the average value, while the (red) bars show the standard
deviation. (b,c) The spin-resolved transmission coefficients for Co majority-spin (spin-up) electrons (b) and
for Co minority-spin (spin-down) electrons as a function of the contact conductance. The lines indicate the
average values and the bars to the standard deviations. (d-f) The same as in panels (a-c) but for Fe contacts.}
\label{fig-Co-Fe}
\end{figure*}

\section{Summary} \label{sec-Conclusions}

To summarize, we have presented a comprehensive experimental and theoretical study of the conductance and 
shot noise in ferromagnetic atomic contacts made of Ni, Co, and Fe, and we have compared the results with 
those for the nobel metal Cu. Our experimental results reveal clear differences between the ferromagnetic 
contacts and those made of Cu such as the absence of any type of conductance quantization and the larger 
values of the Fano factor for any conductance value (including the saturation value for large contacts) 
in the ferromagnetic case. Our theoretical results, which are able to satisfactorily reproduce our main 
experimental observations, clearly show that the transport properties of the ferromagnetic contacts can 
be explained in the framework of quantum coherent transport.

Our results demonstrate that the $d$ orbitals (especially those of the minority-spin electrons) play a 
fundamental role in the transport through ferromagnetic atomic contacts. The contribution of these $d$ orbitals 
leads to the appearance of partially-open channels, which explains the absence of conductance quantization 
and the large values of the Fano factor, as compared to noble metals. Moreover, the contribution of 
minority-spin channels makes the ferromagnetic contacts more conductive than the noble metallic contacts and 
leads to negative values of the spin polarization, both observations in stark contrast 
with the behavior of macroscopic metallic wires. Thus this work provides a textbook example of how the 
transport properties of metallic wires can drastically change upon shrinking their characteristic dimensions 
to the atomic scale, a change that is due to modification in the dominant transport mechanism.

\section{Acknowledgements}

The authors thank A.~Halbritter for fruitful discussions.
M.M.\ and P.N.\ acknowledge financial support from the SFB767 and computer time granting from the NIC and 
the bwHPC framework program of the State of Baden-W\"urttemberg. J.C.C.\ acknowledges financial support from 
the Spanish Ministry of Economy and Competitiveness (Contract No.\ FIS2014-53488-P). O.T.\ thanks the Harold 
Perlman family for their support and acknowledges funding by the Israel Science Foundation and the Minerva 
Foundation.

\appendix

\section{Some additional theoretical results for Co and Fe contacts}

\begin{figure}[t]
\begin{center} \includegraphics[width=0.9\columnwidth,clip]{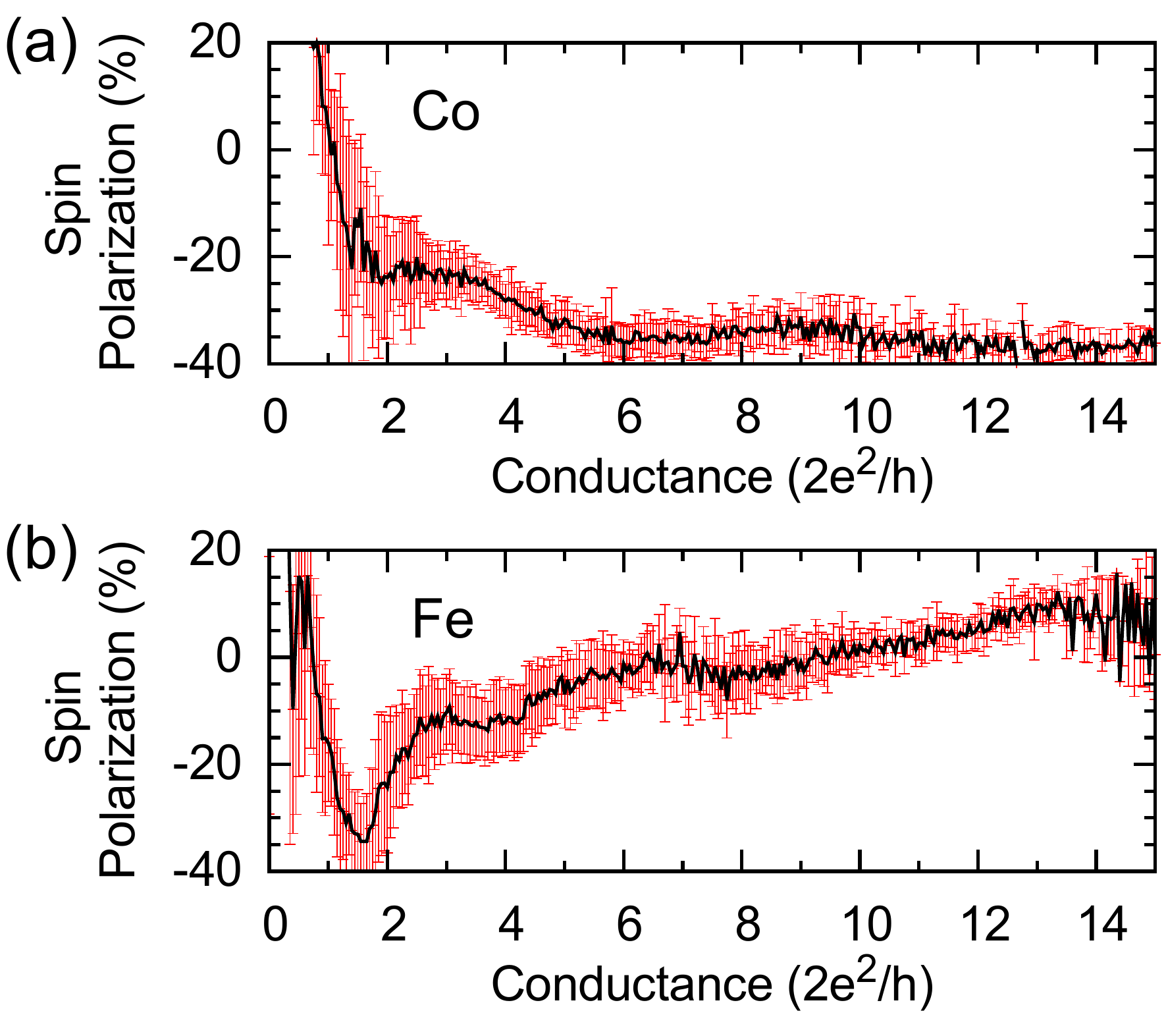} \end{center}
\caption{(Color online) Computed spin polarization of the current as a function of the conductance for
Co (a) and Fe (b). The solid lines correspond to the average value and the bars to the standard deviation.}
\label{fig-spinpol-Co-Fe}
\end{figure}

For completeness, we display in Fig.~\ref{fig-Co-Fe} our theoretical results for the Fano factor and the
spin-resolved channel distributions for the Co and Fe contacts. As explained in the main text, these results
are qualitatively similar to those of Ni and confirm the general conclusions drawn on the nature of the
transport properties of ferromagnetic atomic contacts. We also show in Fig.~\ref{fig-spinpol-Co-Fe}(a,b) 
the corresponding results for the current spin polarization in Co and Fe contacts. Notice that the Co case is 
very similar to the Ni (Fig.~\ref{fig-spinpol}), while in the Fe case the current spin polarization exhibits 
a sign change as a function of the contact size and it adopts rather small values for large contacts. An 
analysis of the channel distribution for Fe, see Fig.~\ref{fig-Co-Fe}(e,f), suggests that this behavior is 
due to a reduced contribution of the minority-spin $d$ bands, as compared with the other two ferromagnetic
metals. Another feature in these results that is worth remarking is the fact that, as in the Ni case, 
the fluctuations in the current polarization are particularly large in the tunnel regime ($G < 1G_0$),
when the contacts are already broken.

%



\begin{thebibliography}{00}

\bibitem{Agrait2003}
N. Agra\"{\i}t, A. Levy Yeyati, and J.~M. van Ruitenbeek, Phys. Rep. {\bf 377}, 81 (2003).

\bibitem{Scheer1998}
E. Scheer, N. Agra\"{\i}t, J.~C. Cuevas, A. Levy Yeyati, B. Ludoph, A. Mart\'{\i}n-Rodero,
G. Rubio, J.~M. van Ruitenbeek and C. Urbina, Nature (London) {\bf 394}, 154 (1998).

\bibitem{vandenBrom1999}
H.~E. van den Brom and J.~M. van Ruitenbeek, Phys. Rev. Lett. {\bf 82}, 1526 (1999).

\bibitem{Wheeler2010}
P.~J. Wheeler, J. Russom, K. Evans, N. King, and D. Natelson, Nano Lett. {\bf 10}, 1287 (2010).

\bibitem{Chen2012}
R. Chen, P.~J. Wheeler, and D. Natelson, Phys. Rev. B {\bf 85} 235455 (2012).

\bibitem{Kumar2012}
M. Kumar, R. Avriller, A. Levy Yeyati, and J.~M. van Ruitenbeek, Phys. Rev. Lett. {\bf 108},
146602 (2012).

\bibitem{Vardimon2013}
R. Vardimon, M. Klionsky, and O. Tal, Phys. Rev. B {\bf 88}, 161404 (2013).

\bibitem{Chen2014}
R. Chen, M. Matt, F. Pauly, P. Nielaba, J.~C. Cuevas, D. Natelson, J. Phys.: Condens. Matter {\bf 26},
474204 (2014).

\bibitem{Ludoph1999}
B. Ludoph and J.~M. van Ruitenbeek, Phys. Rev. B {\bf 59}, 12290 (1999).

\bibitem{Tsutsui2013}
M. Tsutsui, T. Morikawa, A. Arima, and M. Taniguchi, Sci. Rep. {\bf 3}, 3326 (2013).

\bibitem{Evangeli2015}
C. Evangeli, M. Matt, L. Rinc\'on-Garc\'{\i}a, F. Pauly, P. Nielaba, G. Rubio-Bollinger,
J.~C. Cuevas, N. Agra\"{\i}t, Nano Lett. {\bf 15}, 1006 (2015).

\bibitem{Smit2004}
R. H. M. Smit, C. Untiedt, and J. M. van Ruitenbeek, Nanotechnology {\bf 15}, S472 (2004).

\bibitem{Lee2013}
W. Lee, K. Kim, W. Jeong, L.A. Zotti, F. Pauly, J.~C. Cuevas, and P. Reddy,
Nature (London) {\bf 498}, 209 (2013).

\bibitem{Bolotin2006a}
K.~I. Bolotin, F. Kuemmeth, and D.~C. Ralph, Phys. Rev. Lett. {\bf 97}, 127202 (2006).

\bibitem{Sokolov2007}
A. Sokolov, E.~Y. Tsymbal, J. Redepenning, and B. Doudin, Nat. Nanotechnol. {\bf 2}, 171 (2007).

\bibitem{Neel2009}
N. N\'eel, J. Kr\"oger, and R. Berndt, Phys. Rev. Lett. {\bf 102}, 086805 (2009).

\bibitem{Egle2010}
S. Egle, C. Bacca, H.~F. Pernau, M. H\"ufner, D. Hinzke, U. Nowak, and E. Scheer,
Phys. Rev. B {\bf 81}, 134402 (2010).

\bibitem{Strigl2015}
F. Strigl, C. Espy, M. B\"uckle, E. Scheer, and T. Pietsch,
Nature Comm. {\bf 6}, 6172 (2015).

\bibitem{Vardimon2015}
R. Vardimon, M. Klionsky and O. Tal, Nano Lett. {\bf 15}, 3894 (2015).

\bibitem{Mott1936a}
N.~F. Mott, Proc. R. Soc. A Math. Phys. Eng. Sci. {\bf 153}, 699 (1936).

\bibitem{Mott1936b}
N.~F. Mott and H. Jones, \emph{The Theory of the Properties of Metals and Alloys}
(Oxford University Press, London, 1936).

\bibitem{Jin2015}
Z. Jin, A. Tkach, F. Casper, V. Spetter, H. Grimm, A. Thomas, T. Kampfrath, M. Bonn, M. Kl\"aui,
and D. Turchinovich, Nat. Phys. {\bf 11}, 761 (2015).

\bibitem{Sirvent1996}
C. Sirvent, J.~G. Rodrigo, S. Vieira, L. Jurczyszyn, N. Mingo, and F. Flores,
Phys. Rev. B {\bf  53}, 16086 (1996).

\bibitem{Costa1997}
J.~L. Costa-Kr{\"a}mer, Phys. Rev. B {\bf 55}, R4875, (1997).

\bibitem{Hansen1997}
K. Hansen, E. Laegsgaard, I. Stensgaard, and F. Besenbacher,
Phys. Rev. B {\bf  56}, 2208 (1997).

\bibitem{Ott1998}
F. Ott, S. Barberan, J.~G. Lunney, J.~M.~D. Coey, P. Berthet, A.~M.  de Leon-Guevara,
and A. Revcolevschi, Phys. Rev. B {\bf  58}, 4656 (1998).

\bibitem{Oshima1998}
H. Oshima and K. Miyano, Appl. Phys. Lett. {\bf  73}, 2203 (1998).

\bibitem{Ono1999}
T. Ono, Y. Ooka, H. Miyajima, and Y. Otani, Appl. Phys. Lett. {\bf  75}, 1622 (1999).

\bibitem{Komori1999}
F. Komori and K. Nakatsuji, J. Phys. Soc. Jap. {\bf  68}, 3786 (1999).

\bibitem{Garcia1999}
N. Garc\'{\i}a, M. Mu{\~n}oz, and Y.-W. Zhao, Phys. Rev. Lett. {\bf  82}, 2923 (1999).

\bibitem{Ludoph2000b}
B. Ludoph and J.~M. van Ruitenbeek, Phys. Rev. B {\bf  61}, 2273, (2000).

\bibitem{Yanson2001}
A.~I. Yanson, Ph.D. thesis, Universiteit Leiden, 2001.

\bibitem{Viret2002}
M. Viret, S. Berger, M. Gabureac, F. Ott, D. Olligs, I. Petej, J.~F. Gregg, C. Fermon,
G. Francinet, and G. Le Goff, Phys. Rev. B {\bf  66}, 220401(R) (2002).

\bibitem{Elhoussine2002}
F. Elhoussine, S. M{\'a}t{\'e}fi-Tempfli, A. Encinas, and L. Piraux,
Appl. Phys. Lett. {\bf  81}, 1681 (2002).

\bibitem{Shimizu2002}
M. Shimizu, E. Saitoh, H. Miyajima, and Y. Otani, J. Magn. Magn. Mat. {\bf  239},
243 (2002).

\bibitem{Gillingham2002}
D. Gillingham, I. Linington, and J. Bland, J. Phys.: Condens. Matter {\bf  14}, L567 (2002).

\bibitem{Rodrigues2003}
V. Rodrigues, J. Bettini, P.~C. Silva, and D. Ugarte, Phys. Rev. Lett. {\bf  91}, 96801 (2003).

\bibitem{Gillingham2003a}
D. Gillingham, C. M\"uller, and J. Bland, J. Phys.: Condens. Matter {\bf  15}, L291 (2003).

\bibitem{Gillingham2003b}
D. Gillingham, I. Linington, C. M\"uller, and J. Bland, J. Appl. Phys. {\bf  93}, 7388 (2003).

\bibitem{Untiedt2004}
C. Untiedt, D.~M.~T. Dekker, D. Djukic, and J.~M. van Ruitenbeek, Phys. Rev. B {\bf  69},
081401(R) (2004).

\bibitem{Gabureac2004}
M. Gabureac, M. Viret, F. Ott, and C. Fermon, Phys. Rev. B {\bf  69}, 100401(R) (2004).

\bibitem{Yang2004}
C.-S. Yang, C. Zhang, J. Redepenning, and B. Doudin, Appl. Phys. Lett. {\bf  84}, 2865 (2004).

\bibitem{Costa2005}
J.~L. Costa-Kr{\"a}mer, M. D\'{\i}az, P.~A. Serena, Appl. Phys. A {\bf 81}, 1539 (2005).

\bibitem{Bolotin2006b}
K.~I. Bolotin, F. Kuemmeth, A.~N. Pasupathy, and D.~C. Ralph, Nano Lett. {\bf  6}, 123 (2006).

\bibitem{Keane2006}
Z.~K. Keane, L.~H. Yu, and D. Natelson, Appl. Phys. Lett. {\bf  88}, 062514 (2006).

\bibitem{Viret2006}
M. Viret, M. Gabureac, F. Ott, C. Fermon, C. Barreteau, G. Aut\'es, and R. Guirardo-Lopez,
Eur. Phys. J. B {\bf 51}, 1 (2006).

\bibitem{Calvo2009}
M.~R. Calvo, J. Fern\'andez-Rossier, J.~J. Palacios, D. Jacob, D. Natelson, C. Untiedt,
Nature {\bf 458}, 1150 (2009).

\bibitem{Halbritter2010}
A. Halbritter, P. Makk, Sz. Mackowiak, Sz. Csonka, M. Wawrzyniak, and J. Martinek,
Phys. Rev. Lett. {\bf 105}, 266805 (2010).

\bibitem{Moriguchi2012}
Y. Moriguchi, K. Yamauchi, S. Kurokawa, A. Sakai, Surf. Sci. {\bf 606}, 928 (2012).

\bibitem{vonBieren2013}
A. von Bieren, A.~K. Patra, S. Krzyk, J. Rhensius, R.~M. Reeve, L.~J. Heyderman,
R. Hoffmann-Vogel, and M. Kl\"aui, Phys. Rev. Lett. {\bf 110}, 067203 (2013).

\bibitem{Velev2005}
J. Velev, R.~F. Sabirianov, S.~S. Jaswal, and E.~Y. Tsymbal, Phys. Rev. Lett. {\bf 94}, 127203 (2005).

\bibitem{Hafner2009}
M. H\"afner, J.~K. Viljas, and J.~C. Cuevas, Phys. Rev. B {\bf 79}, 140410(R) (2009).

\bibitem{Shi2007a}
S.-F. Shi and D.~C. Ralph, Nat. Nanotechnol. {\bf 2}, 522 (2007).

\bibitem{Shi2007b}
S.-F. Shi, K.~I. Bolotin, F. Kuemmeth, and D.~C. Ralph, Phys. Rev. B {\bf 76}, 184438 (2007).

\bibitem{Cuevas1998}
J.~C. Cuevas, A. Levy Yeyati and A. Martin-Rodero, Phys. Rev. Lett. {\bf 80}, 1066 (1998).

\bibitem{Martin2001}
A. Mart\'{\i}n-Rodero, A. Levy Yeyati, and J.~C. Cuevas, Physica C {\bf 352}, 67 (2001).

\bibitem{Krstic2002}
S. Krsti\'c, X.-G. Zhang, and W.~H. Butler, Phys. Rev. B {\bf 66}, 205319 (2002).

\bibitem{Smogunov2002}
A. Smogunov, A. Dal Corso, and E. Tossati, Surf. Sci. {\bf 507}, 609 (2002);
{\bf 532}, 549 (2003).

\bibitem{Delin2003}
A. Delin and E. Tosatti, Phys. Rev. B {\bf 68}, 144434 (2003).

\bibitem{Velev2004}
J. Velev and W.~H. Butler, Phys. Rev. B {\bf 69}, 094425 (2004).

\bibitem{Bagrets2004}
A. Bagrets, N. Papanikolaou, and I. Mertig, Phys. Rev. B {\bf 70}, 064410 (2004).

\bibitem{Rocha2004}
A.~R. Rocha and S. Sanvito, Phys. Rev. B {\bf 70}, 094406 (2004).

\bibitem{Jacob2005}
D. Jacob, J. Fern\'andez-Rossier, and J.~J. Palacios, Phys. Rev. B {\bf 71},
220403(R) (2005).

\bibitem{Wierzbowska2005}
M. Wierzbowska, A. Delin, and E. Tosatti, Phys. Rev. B {\bf 72}, 035439 (2005).

\bibitem{Dalgleish2005}
H. Dalgleish and G. Kirczenow, Phys. Rev. B {\bf 72}, 155429 (2005).

\bibitem{Khomyakov2005}
P.~A. Khomyakov, G. Brocks, V. Karpan, M. Zwierzycki, and P.~J. Kelly,
Phys. Rev. B {\bf 72}, 035450 (2005).

\bibitem{Fernandez-Rossier2005}
J. Fern\'andez-Rossier, D. Jacob, C. Untiedt, and J.~J. Palacios,
Phys. Rev. B {\bf 72}, 224418 (2005).

\bibitem{Smogunov2006}
A. Smogunov, A. Dal Corso, and E. Tosatti, Phys. Rev. B {\bf 73}, 075418 (2006).

\bibitem{Pauly2006}
F. Pauly, M. Dreher, J.~K. Viljas, M. H\"afner, J.~C. Cuevas, and P. Nielaba,
Phys. Rev. B {\bf 74}, 235106 (2006).

\bibitem{Jacob2006}
D. Jacob and J.~J. Palacios, Phys. Rev. B {\bf 73}, 075429 (2006).

\bibitem{Autes2006}
G. Aut\'es, C. Barreteau, D. Spanjaard, and M.~C. Desjonqu\`res,
J. Phys.: Condens. Matter {\bf 18}, 6785 (2006).

\bibitem{Xia2006}
K. Xia, M. Zwierzycki, M. Talanana, P.~J. Kelly, and G.~E.~W. Bauer,
Phys. Rev. B {\bf 73}, 064420 (2006).

\bibitem{Rocha2007}
A.~R. Rocha, T. Archer, and S. Sanvito, Phys. Rev. B {\bf 76}, 054435 (2007).

\bibitem{Tung2007}
J.~C. Tung and G.~Y. Guo, Phys. Rev. B {\bf 76}, 094413 (2007).

\bibitem{Bagrets2007}
A. Bagrets, N. Papanikolaou, and I. Mertig, Phys. Rev. B {\bf 75}, 235448 (2007).

\bibitem{Autes2008}
G. Aut\'es, C. Barreteau, M.-C. Desjonqu\`eres, D. Spanjaard, and M. Viret,
Europhys. Lett. {\bf 83}, 17010 (2008).

\bibitem{Jacob2008}
D. Jacob, J. Fern\'andez-Rossier, and J.~J. Palacios, Phys. Rev. B {\bf 77}, 165412 (2008).

\bibitem{Tao2010}
K. Tao, I. Rungger, S. Sanvito, and V.~S. Stepanyuk, Phys. Rev. B {\bf 82}, 085412 (2010).

\bibitem{Hardrat2012}
B. Hardrat, N.-P. Wang, F. Freimuth, Y. Mokrousov, and S. Heinze,
Phys. Rev. B {\bf 85}, 245412 (2012).

\bibitem{Xie2012}
Y.-q. Xie, Q. Li, L. Huang, X. Ye, and S.-H. Ke, Appl. Phys. Lett. {\bf 101}, 192408 (2012).

\bibitem{Calvo2012}
M.R. Calvo, D. Jacob, and C. Untiedt, Phys. Rev. B {\bf 86}, 075447 (2012).

\bibitem{Tan2013}
Z.Y. Tan, X.-l. Zheng, X. Ye, Y.-q. Xie, and S.-H. Ke, Appl. Phys. Lett. {\bf 114}, 063711 (2013).

\bibitem{Scheer1997}
E. Scheer, P. Joyez, D. Esteve, C. Urbina, and M.~H. Devoret, Phys. Rev. Lett. {\bf 78}, 3535 (1997).

\bibitem{Muller1992}
C. J. Muller, J. M. van Ruitenbeek, and L.~J. de Jongh, Phys. Rev. Lett. {\bf 69}, 140 (1992).

\bibitem{Dreher2005}
M. Dreher, F. Pauly, J. Heurich, J.~C. Cuevas, E. Scheer, P. Nielaba,
Phys. Rev. B {\bf 72}, 075435 (2005).

\bibitem{Pauly2011}
F. Pauly, J.~K. Viljas, M. B\"urkle, M. Dreher, P. Nielaba, and J.~C. Cuevas,
Phys. Rev. B {\bf 84}, 195420 (2011).

\bibitem{Schirm2013}
C. Schirm, M. Matt, F. Pauly, J.~C. Cuevas, P. Nielaba, and E. Scheer, Nat. Nanotechnol. {\bf 8}, 645 (2013).

\bibitem{Plimpton1995}
S. Plimpton, J. Comp. Phys. {\bf 117}, 1 (1995).

\bibitem{LAMMPS-web}
http://lammps.sandia.gov

\bibitem{Sheng2011}
H.~W. Sheng, M.~J. Kramer, A. Cadien, T. Fujita, and M.~W. Chen, Phys. Rev. B {\bf 83}, 134118 (2011).

\bibitem{Purja2012}
G.~P. Purja Pun and Y. Mishin, Phys. Rev. B {\bf 86}, 134116 (2012).

\bibitem{Mendelev2003}
M.~I. Mendelev, S. Han, D.~J. Srolovitz, G.~J. Ackland, D.~Y. Sun, and M. Asta,
Phil. Mag. A {\bf 83}, 3977 (2003).

\bibitem{Frenkel2004}
D. Frenkel and  B. Smit, \emph{Understanding Molecular Simulation} (Academic Press, San Diego, 2004).

\bibitem{Mehl1996}
M.~J. Mehl and D.~A. Papaconstantopoulos, Phys. Rev. B {\bf 54}, 4519 (1996).

\bibitem{Mehl1998}
M.~J. Mehl and D.~A. Papaconstantopoulos, \emph{Computational Materials Science},
edited by C. Fong (World Scientific, Singapore, 1998).

\bibitem{Bacalis2001}
N.~C. Bacalis, D.~A. Papaconstantopoulos, M.~J. Mehl, and M. Lach-hab, Physica B {\bf 296}, 125 (2001).

\bibitem{Guinea1983}
F. Guinea, C. Tejedor, F. Flores, and E. Louis, Phys. Rev. B {\bf 28}, 4397 (1983).

\bibitem{Pauly2008}
F. Pauly, J.~K. Viljas, U. Huniar, M. H\"afner, S. Wohlthat, M. B\"urkle, J.~C. Cuevas, G. Sch\"on,
New J. Phys. {\bf 10}, 125019 (2008).

\bibitem{Cuevas2010}
J.~C. Cuevas and E. Scheer, \emph{Molecular Electronics: An Introduction to Theory and Experiment}
(World Scientific, Singapore, 2010).

\bibitem{Rubio1996}
G. Rubio, N. Agrait, S. Vieira, Phys. Rev. Lett. {\bf 76}, 2302 (1996).

\bibitem{Sharvin1965}
Yu.V. Sharvin, Sov. Phys.-JETP {\bf 21}, 655 (1965) [Zh. Eksp. Teor. Fiz. {\bf 48}, 984 (1965)].

\bibitem{Blanter2000}
Y.~M. Blanter and M. B\"uttiker, Phys. Rep. {\bf 336}, 1 (2000).

\bibitem{vandenBrom2000}
H.~E. van den Brom, \emph{Noise Properties of Atomic-Size Contacts}, Universiteit Leiden (2000).

\bibitem{Nagaev1992}
K.~E. Nagaev, Phys. Lett. A {\bf 169}, 103 (1992).

\bibitem{Beenakker1992}
C.~W.~J. Beenakker and M. B\"uttiker, Phys. Rev. B {\bf 46}, 1889 (1992).

\bibitem{Dorokhov1982}
O.~N. Dorokhov, JETP Lett. {\bf 36}, 318 (1982).

\bibitem{Beenakker1997}
C.~W.~J. Beenakker, Rev. Mod. Phys. {\bf 69}, 731 (1997).

\bibitem{Meservey1994}
R. Meservey and P.~M. Tedrow, Phys. Rep. {\bf 238}, 173 (1994).

\bibitem{Burki1999}
J. B\"urki, C.~A. Stafford, X. Zotos, and D. Baeriswyl, Phys. Rev. B {\bf 60}, 5000 (1999).

\bibitem{Burki1999b}
J. B\"urki and C.~A. Stafford, Phys. Rev. Lett. {\bf 83}, 3342 (1999).

\bibitem{Beenakker1994}
C.~W.~J. Beenakker and J.~A. Melsen, Phys. Rev. B {\bf 50}, 2450 (1994).

\bibitem{Riquelme2005}
J.~J. Riquelme, L. de la Vega, A. Levy Yeyati, N. Agra\"{\i}t, A. Martin-Rodero,
and G. Rubio-Bollinger, Europhys. Lett. {\bf 70}, 663 (2005).

\bibitem{Ludoph2000c}
B. Ludoph and J.~M. van Ruitenbeek, Phys. Rev. B {\bf 61}, 2273 (2000).

\bibitem{Erts2000}
D. Erts, H. Olin, L. Ryen, E. Olsson, and A. Th\"ol\'en, Phys. Rev. B {\bf 61}, 12725 (2000).

\bibitem{deJong1995}
M.~J.~M. de Jong and C.~W.~J. Beenakker, Phys. Rev. B {\bf 51}, 16867 (1995).

\end{thebibliography}
\end{document}